\documentclass[12pt,english]{article}
\pdfoutput=1
\usepackage[T1]{fontenc}
\usepackage{amssymb}
\usepackage{amsmath}
\usepackage{cite}
\usepackage{epsfig}
\usepackage{scalerel}
\usepackage[unicode=true,bookmarks=true,bookmarksnumbered=false,bookmarksopen=false,
 breaklinks=false,pdfborder={0 0 1},backref=false,colorlinks=true]
 {hyperref}
\usepackage[hang,flushmargin]{footmisc}
\usepackage{color}
\usepackage{shuffle}
\usepackage{cleveref}
\usepackage{nicematrix}

\usepackage[latin1]{inputenc}
\usepackage{amsmath}
\usepackage{amsfonts}
\usepackage{amssymb}

\usepackage{graphicx}
\usepackage[export]{adjustbox}

\usepackage[most]{tcolorbox}

\usepackage{bm}
\newcommand{\vv}[1]{{#1}}

\usepackage{tikz,environ}
\usetikzlibrary{decorations}
\usetikzlibrary{arrows}
\usetikzlibrary{decorations.markings}
\tikzset{
	hard/.style={postaction={decorate},
		line width=0.5mm
	},
		soft/.style={postaction={decorate},
		line width=0.5mm, dashed
	},
	momentum/.style={postaction={decorate},
	line width=0.5mm,
	color=gray,
	decoration={
    markings,
    mark=at position 0.8 with {\arrow{stealth}}}
    },
	hardarrow/.style={postaction={decorate},
	line width=0.5mm,
	decoration={
    markings,
    mark=at position 0.6 with {\arrow{stealth}}}},
}

\usepackage[bbgreekl]{mathbbol}
\DeclareMathSymbol\bbDelta  \mathord{bbold}{"01}

\makeatletter
\def\simgt{\mathrel{\lower2.5pt\vbox{\lineskip=0pt\baselineskip=0pt
           \hbox{$>$}\hbox{$\sim$}}}}
\def\simlt{\mathrel{\lower2.5pt\vbox{\lineskip=0pt\baselineskip=0pt
           \hbox{$<$}\hbox{$\sim$}}}}
\makeatother

\numberwithin{equation}{section}
\newcommand{\be}{\begin{equation}}
\newcommand{\ee}{\end{equation}}
\newcommand{\bea}{\begin{eqnarray}}
\newcommand{\eea}{\end{eqnarray}}
\newcommand{\eq}[2]{\be\begin{aligned}#1 \label{#2}\end{aligned}\ee}

\Crefname{equation}{Eq.}{Eqs.}

\ifdefined \Ref
\renewcommand{\Ref}[1]{Ref.~\cite{#1}}
\else
\newcommand{\Ref}[1]{Ref.~\cite{#1}}
\fi

\newcommand{\Fig}[1]{Fig.~\ref{#1}}
\newcommand{\Eq}[1]{Eq.~\eqref{#1}}
\newcommand{\Eqs}[1]{\Cref{#1}}
\newcommand{\Sec}[1]{Sec.~\ref{#1}}

\newcommand{\App}[1]{App.~\ref{#1}}

\newcommand{\bra}[1]{\langle #1 |}
\newcommand{\ket}[1]{| #1 \rangle}

\allowdisplaybreaks
\hypersetup{citecolor=blue,linkcolor=blue,urlcolor=blue}
\begin{document}

\baselineskip=18pt

\hfill   CALT-TH-2023-002

\vspace{2.5cm}
\thispagestyle{empty}
\begin{center}
{\LARGE \bf
Soft Phonon Theorems
}
 \end{center}
 \bigskip
\begin{center}{\large Clifford Cheung, Maria Derda, Andreas Helset, and Julio Parra-Martinez }\end{center}
\begin{center}
\it Walter Burke Institute for Theoretical Physics,\\[-1mm]
California Institute of Technology, Pasadena, California 91125\\[1.5mm]
\end{center}

\bigskip
\centerline{\large\bf Abstract}

\begin{quote} \small

A variety of condensed matter systems describe gapless modes that can be interpreted as Nambu-Goldstone bosons of spontaneously broken Poincar\'e symmetry.  In this paper we derive new soft theorems constraining the tree-level scattering of these degrees of freedom, as exhibited in solids, fluids, superfluids, and framids.  These soft theorems are in one-to-one correspondence with various broken symmetries, including spacetime translations, Lorentz boosts, and, for the case of fluids, volume-preserving diffeomorphisms. We also implement a bootstrap in which the enhanced vanishing of amplitudes in the soft limit is taken as an input, thus sculpting out a subclass of exceptional solid, fluid, and framid theories.

\end{quote}

\setcounter{footnote}{0}

\newpage

\setcounter{tocdepth}{2}

\tableofcontents    

\newpage

\section{Introduction}

The seminal work of Nambu and Goldstone  \cite{Nambu:1960tm,Goldstone:1961eq,Goldstone:1962es} revealed a deep connection between spontaneously broken internal symmetries and a corresponding set of gapless  degrees of freedom.  These Nambu-Goldstone bosons (NGBs) parameterize a continuous degeneracy of vacua and transform nonlinearly under the broken symmetries. Notably, spontaneous symmetry breaking often mandates universal features in scattering, as perhaps best illustrated by the Adler zero \cite{Adler:1964um}, which refers to the vanishing of certain NGB amplitudes in the soft limit.

As is well-known, similar statements apply to the spontaneous breaking of {\it spacetime symmetries}\cite{Leutwyler:1996er,Leutwyler:1993gf, Son:2002zn, Andersen:2002nd, Brauner:2010wm}, albeit with a fewer number of NGBs than naively expected \cite{Ivanov:1975zq,Low:2001bw, Watanabe:2013iia, Brauner:2014aha}. More recently, it has also been suggested that spontaneous breaking of Poincar\'e invariance is not merely a calling card of certain condensed matter systems, but can actually be elevated to an {\it organizing principle} for these theories \cite{Nicolis:2015sra}.  In this approach, nonrelativistic effective field theories (EFTs) are classified by their spacetime symmetry breaking pattern, yielding a rich array of physical systems corresponding to the phonon excitations at zero temperature of perfect solids and fluids, as well as modes of a superfluid.  The authors of \cite{Nicolis:2015sra} also discovered some exotic, yet-to-be-experimentally-realized systems which include the framid, whose corresponding framon degree of freedom exhibits the minimal nonlinear realization of spontaneous Lorentz symmetry breaking.

In this paper we study the soft behavior of scattering amplitudes of NGBs arising from the spontaneous breaking of spacetime symmetries. Our analysis focuses on the nonrelativistic EFTs classified in \cite{Nicolis:2015sra} by the symmetry breaking pattern corresponding to solids, fluids, superfluids, and framids. In all of these systems, the group of spatial rotations is preserved at low energies, so the NGBs reside in a linear representation of $SO(3)$.  For example, the superfluid phonon is described by a scalar field $\pi$, while solid and fluid phonons and framons are described by a three-vector field $ \pi^i$.\footnote{Throughout, we use Greek letters $\mu,\nu,\rho,\cdots,$ to denote four-vector indices, late Latin letters $i,j,k,\cdots,$ to denote three-vector indices, and early Latin letters $a,b,c,\cdots,$ to denote external particle labels. For products of three-momenta we will sometimes employ the shorthand, $\vv p \cdot \vv q = \vv p^i \vv q_i$.} 
In general, the latter NGBs nonlinearly realize the underlying broken spacetime symmetries via
\eq{  \pi^i \rightarrow  \pi^i+ \alpha^i + \beta^i_{\,\,\,j}  \pi^j+\cdots,
}{}
for parameters $\alpha$ and $\beta$, which are in general {\it position-dependent}, and where the ellipses denote terms higher order in the field.   In our analysis, we focus primarily on the case in which the broken spacetime symmetry generators are translations or boosts, but also consider the case of broken spatial diffeomorphisms in a fluid.

Following the logic of \cite{Green:2022slj}, we use current conservation to derive a broad class of soft theorems applicable to NGBs arising from the spontaneous breaking of {\it any} symmetry.  Technically, our results apply to any symmetry breaking pattern involving spacetime or internal symmetries or both.\footnote{While the present work focuses solely on theories which preserve $SO(3)$ rotation symmetry, this is actually not required for our soft theorem.   In particular, our results apply to any symmetry breaking pattern that preserves some version of spacetime translations in the broken phase such that energy and momentum are well-defined.  We leave an analysis of more drastic symmetry breaking patterns for later work.} 
The schematic form of our soft theorem is 
\eq{
\lim_{q\rightarrow 0}\alpha [ A_{n+1} ] &\sim -  \lim_{q\rightarrow 0}\alpha\left[ \sum V_3 \Delta A_n\right]- \sum \beta  [A_n] \, ,
}{soft_theorem_schematic}
where the soft limit $q\rightarrow 0$ corresponds to sending the energy and momentum components of the soft leg to zero.  Here both sides of \Eq{soft_theorem_schematic} are ${\cal O}(q^0)$ in the soft momentum, which is to say that terms ${\cal O}(q^{1})$ or higher have been dropped.    The summations above run over all external legs in $A_n$, which are assumed to be hard.

Since $\alpha$ and $\beta$ are in general functions of spacetime, they act as {\it differential operators} on the external momenta.  In particular, in \Eq{soft_theorem_schematic}, $\alpha$ acts on the soft energy or momentum, while $\beta$ acts on the energy or momenta of each hard leg in $A_n$.  
Depending on the differential degree of $\alpha$ and $\beta$, they will extract different powers in the soft expansion of the amplitudes.  For example, if $\alpha$ is a constant, then \Eq{soft_theorem_schematic} extracts the ${\cal O}(q^0)$ piece of $A_{n+1}$, while if $\alpha$ is a single derivative with respect to the soft energy or momentum, then it probes the ${\cal O}(q^1)$ piece of $A_{n+1}$.

\Eq{soft_theorem_schematic} is an on-shell soft theorem because both the left- and right-hand sides are operations acting on the on-shell amplitudes $A_{n+1}$ and $A_n$.
Furthermore, as required of any physical on-shell scattering, the soft theorem in \Eq{soft_theorem_schematic} is satisfied irrespective of the choice of field basis.  This feature is actually rather miraculous when one considers that \Eq{soft_theorem_schematic} depends explicitly on the {\it off-shell} three-point vertex $V_3$, which is field basis dependent along with the symmetry parameter $\beta$.  However, as we will later argue, any change of field basis that sends $V_3$ and $\beta$ to an alternative choice of $V_3'$ and $\beta'$ {\it necessarily} cancels in the soft theorem. The fact that the soft theorem is on-shell makes our results distinct from the soft theorems for correlation functions derived in \cite{Assassi:2012zq,Finelli:2017fml,Hui:2018cag,Jazayeri:2019nbi,Hui:2022dnm}.

The structure of this paper is as follows.  In \Sec{sec:softTheorem} we state the general soft theorem and present a proof as well as a discussion of its invariance under changes of field basis. We then turn to concrete examples of theories that satisfy the soft theorem: superfluids in \Sec{sec:Superfluids}, solids in \Sec{sec:Solids}, fluids in \Sec{sec:Fluids}, and framids in \Sec{sec:Framids}. In \Sec{sec:bootstrap} we discuss a soft bootstrap for nonrelativistic theories, and we conclude in \Sec{sec:Conclusion}.

\section{Soft Theorem}
\label{sec:softTheorem}

\subsection{Degrees of Freedom}

In this work we focus on nonrelativistic theories describing a NGB arising from a spontaneously broken spacetime symmetry.\footnote{For simplicity, we focus on the case of NGBs of type I with a linear dispersion relation. While we do not explicitly analyze NGBs of type II with quadratic dispersion relations (see \cite{Lange:1965zz, Lange:1966zz, Nielsen:1975hm, Watanabe:2011dk, Watanabe:2011ec, Watanabe:2012hr, Kapustin:2012cr, Watanabe:2013iia, Watanabe:2014fva}), all of our results, including the general soft theorem, should apply more generally.} For concreteness, let us consider here the case of an $SO(3)$ vector field $ \pi^i$,
which transforms linearly under
\eq{
\textrm{spatial rotations:} \quad  \pi^i \rightarrow  
R^i_{\,\,\,j}  \pi^j \, ,
}{}
for a constant orthogonal matrix $R$.   Here we emphasize that $ \pi^i$ does {\it not} describe a gauge theory in the conventional sense, since all three of its components are physical: they correspond to one longitudinal and two transverse modes of the NGB, which we describe in terms of one-particle states, $|\omega, \vv p, L\rangle$ and $|\omega, \vv p, T\rangle$, respectively.   These states overlap with the $ \pi^i$ field according to
\eq{
\bra{0}   \pi^i(t,x) \ket{\omega, \vv p, L}
&=  \vv e_{L}^i(\vv p) e^{-i \omega t+i p\cdot x} \,,\\
\bra{0}   \pi^i(t,x) \ket{\omega, \vv p, T}
&=  \vv e_{T}^i(\vv p) e^{-i \omega t+i p\cdot x} 
\, .
}{overlap}
Here $\omega$ and $\vv p$ are the energy and three-momentum of the particle.  Depending on the particle type, these quantities obey the dispersion relations,
\eq{
 \omega^2 - c_L^2 \vv p^2 =0\qquad \textrm{or} \qquad  \omega^2 - c_T^2 \vv p^2=0 \, ,
}{OS-dispersion}
where $c_L$ and $c_T$ are the speeds of sound for the longitudinal and transverse modes, respectively.  The polarization vectors in \Eq{overlap} satisfy 
\eq{
\vv p_i \vv e^i_T = 0 \qquad \textrm{and} \qquad \epsilon_{ijk} \vv p^j \vv e_{L}^k=0 \, .
}{OS-pol} 
This implies that $\vv e_L $ encodes the single longitudinal mode while $\vv e_T$ encodes the two transverse modes.  For explicit calculations, we will use the unit normalized longitudinal polarization,
\eq{
e^i_L = \frac{c_L p^i}{\omega} \, .
}{}
We can think of \Eq{OS-dispersion} and \Eq{OS-pol} as the on-shell conditions for the kinematic variables that characterize the phonon modes.
 
 Here it will be convenient to define the projection operators,
 \eq{
\Pi^{i j}_L (p)&=\frac{\vv p^{i}\vv p^{j}}{\vv p^2} \,, \\
\Pi^{i j}_T(p) &=\delta^{i j}-\frac{\vv p^{i}\vv p^{j}}{\vv p^2} \,,
}{}
which leave the polarizations invariant, so
\eq{
\Pi^{ij}_L (p) \vv e_{Lj} =  \vv e_L^i \qquad \textrm{and} \qquad
\Pi^{ij}_T(p) \vv e_{Tj} =  \vv e_T^i \,.
}{}
In terms of these projectors, the phonon propagator is
\eq{
\Delta^{i j}(\omega, p) &= \frac{\Pi^{i j}_L(p)}{\omega^2 - c_L^2 \vv p^2} + \frac{\Pi^{i j}_T(p)}{\omega^2 - c_T^2 \vv p^2}\,,
}{propagator}
whose inverse, $\Delta^{-1}_{ij}(\omega,p)$, is the two-point Lagrangian term in momentum space.

To define the $n$-point scattering amplitude of phonons we define a set of external particles labelled by $a=1, \cdots, n$, whose corresponding momenta are $p_{a}^\mu = (\omega_a, \vv p^i_a)$, with speed of sound $c_a$ and polarization vector $e_a^i$ which is chosen to be either longitudinal or transverse.  The $n$-point scattering amplitude is 
\eq{
A^{i_1 \cdots i_n }_n(p_1,\cdots, p_n) = \langle 0| \left(| \omega_1, p_1 \rangle^{i_1} \cdots | \omega_n, p_n \rangle^{i_n} \right)  \,,
}{project_onto_onshell}
where we have defined a shorthand for a state $| \omega, p \rangle^{i}$ that carries an arbitrary polarization and is related to the physical longitudinal and transverse states defined previously by $|\omega, p,L\rangle =  e_L^{i} | \omega, p \rangle_{i}$ and $|\omega, p,T\rangle =  e_T^{i} | \omega, p \rangle_{i}$, respectively.  

The quantity $A^{i_1 \cdots i_n }_n$ is simply the amputated correlation function of phonon fields, which can be computed straightforwardly using Feynman diagrams.  To compute the physical amplitude we simply dot this object into external polarization vectors.  Note that we have used a schematic notation in which $A^{i_1 \cdots i_n }_n$ is written as a function of just the three-momenta $p_1,\cdots, p_n$.  However, since we are interested in on-shell kinematics, momenta and energies can be interchanged freely.  So in explicit calculations, our actual amplitudes may be functions of energies as well as the three-momenta.

\subsection{Proof of Theorem}

To begin, recall that the NGB of spontaneous spacetime symmetry breaking transforms nonlinearly under the broken symmetry transformations,
\eq{
\pi^i \rightarrow \pi^i + \delta \pi^i \, .
}{}
In general, the nonlinearly realized symmetry transformation may also involve changes of coordinates, but this will not be important for our analysis. In addition, we will assume that the Lagrangian does not depend on the second and higher derivatives of fields.
The statement that the Lagrangian is invariant implies that
\eq{
L \rightarrow L + \delta L \, ,
}{}
where the Lagrangian variation is
\eq{
\delta L = \partial_\mu \left( \delta \pi^i \frac{\delta L}{ \delta \partial_\mu \pi^i}  \right) - \delta \pi^i E_i = \partial_\mu K^\mu \, .
}{is_symmetry}
Here $K$ describes any shift of the Lagrangian by a total derivative, as would often appear in a spacetime symmetry transformation.  Meanwhile, $E$ denotes the equation of motion,
\eq{
E_i =  \partial_\mu \left( \frac{\delta L}{ \delta \partial_\mu \pi^i} \right)- \frac{\delta L}{\delta \pi^i} \overset{\textrm{\tiny on-shell}}{=} 0 \, ,
}{}
which vanishes on the support of on-shell, physical field configurations.  Recalling the definition of the conserved current,
\eq{
J^\mu = \delta \pi^i \frac{\delta L}{ \delta \partial_\mu \pi^i}- K^\mu \, ,
}{}
which satisfies the conservation equation,
\eq{
\partial_\mu J^\mu = \delta \pi^i E_i \overset{\textrm{\tiny on-shell}}{=} 0 \, ,
}{current_conservation}
for on-shell configurations of fields.  In order to derive our soft theorem we evaluate matrix elements of the above equation, keeping contributions up to ${\cal O}(q^1)$ in the soft limit.

For later convenience, let us define a bracket that acts on a local field operator ${\cal O}(t,x)$ via
\eq{
\langle {\cal O} \rangle \delta^4(p_1 +\cdots + p_n) = \lim_{q\rightarrow 0} \int dt d^3x\,  e^{-i\omega t} e^{i q\cdot x} \langle 0 | {\cal O}(t,x) \left( | \omega_1, p_1\rangle^{i_1}  \cdots | \omega_n, p_n\rangle^{i_n}  \right) \, ,
}{}
which is the matrix element obtained by sandwiching the operator between a set of on-shell physical states with arbitrary polarizations.
 By construction, the field operator itself is imparted with energy $\omega$ and three-momentum $q$ which are taken to zero, yielding a soft limit.    Throughout, we assume that an on-shell momentum flows through the operator, so $\omega$ also scales as $q$ and is implicitly sent to zero in the soft limit.
 
To derive a soft phonon theorem we evaluate \Eq{current_conservation} as an operator equation sandwiched between on-shell physical states. To this end, let us define a general parameterization of the infinitesimal shift of the NGB field,
\eq{
\delta \pi^i = \alpha^i + \beta^i_{\,\,\,j} \pi^j +\cdots \, ,
}{delta_pi}
where $\alpha$ and $\beta$ are spacetime-dependent in general and the ellipses denote terms that are higher order in the field.  Meanwhile, the equation of motion takes the general form,  
\eq{
E_i = V_{2ij} \pi^j + \tfrac12 V_{3ijk} \pi^j \pi^k +\cdots \, ,
}{EOM}
where $V_{2ij}$ and $V_{3ijk}$ correspond to the two- and three-point Lagrangian terms.  Going to momentum space, the Feynman propagator and three-point Feynman vertex are equal to $\Delta_{ij}=V_{2ij}^{-1}$ and $V_{3ijk}$, all multiplied by $i$.

\begin{figure}[t]
    \centering
    \begin{tikzpicture}
            \coordinate (d) at (-3.75, 0);
            \coordinate (c) at (-3, 0);	
	    	\coordinate (a) at (-2, 1);
	        \node[left]  at (c) {$a$};
	    	
	    	\coordinate (f) at (1, 1);
	    	\coordinate (l) at (1, -1);
	    	
	    	\coordinate (m4) at (-2,0);
	    	
	    	\coordinate (mn) at (-0.25, 0);
	    
	    	\coordinate (d1) at (1.00, 0);
	    	\coordinate (d2) at (0.85, 0.5);
	    	\coordinate (d3) at (0.85, -0.5);
	    	
	    	\draw [hard] (c) -- (m4);
	    	\draw [hard] (mn) -- (m4);
	    	
	    	\draw [hard] (f) -- (mn);
	        \draw [hard] (l) -- (mn);
	    	
            \draw[fill=lightgray, opacity=1] (mn) circle (0.65);
            	\node at (mn) {$\mathcal A_{n}$};
            
            \draw[fill=white, opacity=1] (m4) circle (0.2);
            \node[below]  at (-2,-0.2) {${\cal O}(q) $};
	    	
            \draw[fill=black, opacity=1] (d1) circle (0.02);
            \draw[fill=black, opacity=1] (d2) circle (0.02);
            \draw[fill=black, opacity=1] (d3) circle (0.02);
\end{tikzpicture}
    \caption{Diagrams computing the contribution from operator insertions on the external legs.}
    \label{fig: current insertions}
\end{figure}
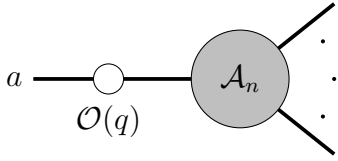

The classical conservation of the current in \Eq{current_conservation} uplifts to the operator statement,
\eq{
0=\langle \partial_\mu J^\mu \rangle = \langle \delta \pi^i E_i \rangle = \langle (\alpha^i + \beta^i_{\,\,\,j} \pi^j  +\cdots )((\Delta^{-1})_{ik} \pi^k + \tfrac12 V_{3ikl} \pi^k \pi^l +\cdots)\rangle \, .
}{op_eq}
To derive the soft theorem we must calculate the matrix elements in each term in \Eq{op_eq}.
In principle one should evaluate all possible insertions of each operator, both on internal and external lines. However, many terms can be neglected since we are only interested in terms at ${\cal O}(q^0)$ but not higher.  In particular, since the first equality in \Eq{op_eq} implies that the matrix element is automatically equipped with an overall factor of $q$, any contributions to $\langle J^\mu\rangle$ which are analytic in $q$ will only generate ${\cal O}(q^1)$ contributions to the matrix element.  Conversely, ${\cal O}(q^0)$ contributions only arise from terms in $\langle J^\mu\rangle$ that go as ${\cal O}(q^{-1})$.  Such terms appear due to soft pole contributions from $q$-dependent propagators, which in turn only  arise from insertions of the operator on external legs (see \Fig{fig: current insertions}).  
On the other hand, operator insertions on internal legs and terms of ${\cal O}(\pi^3)$ or higher yield terms that are analytic in $q$ and thus vanish in the $q\rightarrow 0$ soft limit.  Thus we can drop all such terms in the evaluation of the right-hand side of \Eq{op_eq}.

Shuffling around terms in \Eq{op_eq}, we arrive at
\eq{
 \langle \alpha_i  (\Delta^{-1})^{ij} \pi_j \rangle &= - \langle \tfrac12 \alpha_i  V^{ijk}_{3} \pi_j \pi_k + \beta_i^{\,\,\,j} \pi_j (\Delta^{-1})^{ik} \pi_k   \rangle_{\rm ext} + \cdots \, ,
}{pre_soft_theorem}
where the ellipses denote irrelevant ${\cal O}(\pi^3)$ contributions and the ``ext'' subscript instructs that the matrix element should be evaluated with the operator inserted only on external legs.   As described above, all insertions of the operator on internal legs are subleading in the soft limit and can be neglected. Importantly, each of the terms in \Eq{pre_soft_theorem} can be recast in terms of on-shell scattering amplitudes.  

As a warmup, let us consider the matrix element
\eq{
\langle  (\Delta^{-1})^{ij} \pi_j \rangle =- \lim_{q\rightarrow 0} A^{i_1 \cdots i_n i}_{n+1} (p_1, \cdots, p_n, q) \,,
}{Delta_inv_pi}
where as before, we use an abbreviated notation where we write out explicitly the three-momentum dependence of functions, but implicitly there is also energy dependence everywhere.
\Eq{Delta_inv_pi} simply says that the matrix element of the one-point function of an amputated field is precisely the amputated $(n+1)$-point amplitude.  In the case where the operator includes the spacetime-dependent factor $\alpha$, we obtain
\eq{
\langle \alpha_i  (\Delta^{-1})^{ij} \pi_j \rangle = -\lim_{q\rightarrow 0} \alpha_i(i \tfrac{\partial}{\partial \omega}, -i\tfrac{\partial}{\partial q}) \left[ A^{i_1 \cdots i_n i}_{n+1} (p_1, \cdots, p_n, q)\right] \, .
}{}
Note that in transforming to momentum space, the dependence of $\alpha$ on time $t$ and position $x$ becomes dependence on $i\tfrac{\partial}{\partial \omega}$ and $-i\tfrac{\partial}{\partial q}$, respectively.  

Meanwhile, the terms involving $\beta$ are written in terms of amplitudes as
\eq{
\langle \beta_i^{\,\,\,j} \pi_j (\Delta^{-1})^{ik} \pi_k   \rangle_{\rm ext}  &= -\sum_{a=1}^n \beta^{i_a}_{\,\,\,\,\, j_a}(i \tfrac{\partial}{\partial \omega_a},-i\tfrac{\partial}{\partial p_a})  \left[A^{i_1 \cdots j_a \cdots i_n }_{n} (p_1, \cdots, p_a, \cdots, p_n)\right] 
\, ,
}{beta_pi_pi}
which corresponds to the sum over $\beta$ acting on each external leg in the $n$-point amplitude. Last but not least, the term $ \langle \tfrac12 \alpha_i  V^{ijk}_{3} \pi_j \pi_k\rangle_{\rm ext}$ is
\begin{align}
&  - \lim_{q\rightarrow 0} \sum_{a=1}^n  \alpha_i(i \tfrac{\partial}{\partial \omega}, -i\tfrac{\partial}{\partial q})  \left[  V^{i\,  i_a}_{3\,\,\,\,\,\,\,j_a} (q,p_a)\Delta^{j_a}_{\,\,\, \,\,k_a}(p_a +q) A^{i_1 \cdots k_a \cdots i_n }_{n}(p_1, \cdots, p_a+q, \cdots, p_n)\right] 
\, .
\label{alpha_V_pi_pi}
\end{align}
Here the propagator and $n$-point amplitude on the right-hand side are evaluated at {\it shifted} external energy and three-momentum, $\omega_a + \omega$ and $ p_a + q$, where we have suppressed the dependence on the former in the various expressions for ease of notation.  At low orders in the soft expansion we can express this alternatively as $(1+ \omega  \tfrac{\partial}{\partial \omega_a} +q^i \tfrac{\partial}{\partial p_a^i})$ acting on these objects.  

As noted earlier, the $n$-point  amplitude is in general a function of three-momenta as well as energies---which is expected since these are generally interchangeable due to the on-shell conditions.  Thus, the energies and three-momenta in the $n$-point amplitude should both be shifted.
 We will discuss later on how this shift is explicitly implemented in order to maintain the on-shell conditions.

In conclusion, each term in \Eq{pre_soft_theorem} can be expressed in terms of differential operators acting on the $(n+1)$-point and $n$-point scattering amplitudes. Hence, \Eq{pre_soft_theorem} implies that
\begin{align}
&\lim_{q\rightarrow 0} \alpha_i(i \tfrac{\partial}{\partial \omega}, -i\tfrac{\partial}{\partial q})   \left[ A^{i_1 \cdots i_n i}_{n+1}(p_1, \cdots, p_n, q) \right] = \nonumber \\
 &\qquad\qquad -\lim_{q\rightarrow 0} \sum_{a=1}^n  \alpha_i(i \tfrac{\partial}{\partial \omega}, -i\tfrac{\partial}{\partial q})  \left[  V^{i\,  i_a}_{3\,\,\,\,\,\,\,j_a}(q,p_a) \Delta^{j_a}_{\,\,\, \,\,k_a}(p_a +q) A^{i_1 \cdots k_a \cdots i_n }_{n}(p_1, \cdots, p_a+q, \cdots, p_n)\right] \nonumber \\
&\qquad\qquad -\sum_{a=1}^n \beta^{i_a}_{\,\,\,\,\, j_a}(i \tfrac{\partial}{\partial \omega_a},-i\tfrac{\partial}{\partial p_a})  \left[ A^{i_1 \cdots j_a \cdots i_n }_{n} (p_1, \cdots, p_a, \cdots, p_n)\right] \, ,
\label{soft_theorem_exact}
\end{align}
which is our final expression for the soft theorem after dropping terms ${\cal O}(q^1)$ or higher.

Let us comment on several subtle aspects of \Eq{soft_theorem_exact}.  First of all, the limit $q\rightarrow 0$ with all other momenta unchanged will not, in general, preserve the on-shell conditions.  Hence, a strict soft limit of this kind is not actually well-defined. The same is true for the shift of energy and momenta on the right-hand side of \Eq{soft_theorem_exact}.   For these reasons, all of the amplitudes in \Eq{soft_theorem_exact} should be evaluated in a minimal basis of kinematic invariants, which we will describe in great detail in Appendix~\ref{sec:kinematics}.\footnote{Throughout, we assume complex kinematics, as commonly used in the study of gauge theory amplitudes.} With this prescription, the amplitudes will be on-shell for any value of $q$ and any value of $p_a$, so the soft limit and the shift of momentum are both well-defined on-shell operations.

\subsection{Field Basis Independence}
\label{sec:FR}

Next, we show how the soft theorem in \Eq{soft_theorem_exact} is invariant under changes of field basis.  To begin, we clarify that there are actually two physically distinct senses in which a soft theorem can be considered field basis invariant.  

The first sense is simply the statement that the soft theorem is valid irrespective of which field basis the quantities $\beta$ and $V_3$ are defined in, which enter explicitly into \Eq{soft_theorem_exact}. If we transform to a different field basis, these quantities will change to $\beta'$ and $V_3'$.  But crucially, all of the manipulations in the previous section still hold.  Hence the soft theorem will still apply, so its validity is field basis independent.

The second sense is more nontrivial, and it is the statement that the on-shell amplitudes $A_{n+1}$ and $A_n$ can be computed in different field bases and the soft theorems will still be satisfied. As is well-known, changes of field basis induce new terms which always {\it vanish on-shell} in the amplitudes.  However, since our soft theorems involve differential operators in energy and momentum, one can worry whether these vanishing terms end up contributing to the soft theorems anyway.  We consider this possibility now, and show that such terms have no effect.

Concretely, we will now show how terms that vanish due to on-shell conditions in $A^{i_1 \cdots i_n }_n$ will always cancel automatically in the soft theorem in \Eq{soft_theorem_exact}. Thus, the soft theorem in \Eq{soft_theorem_exact} commutes with the on-shell condition, ensuring the field basis independence of the soft theorem. We assume $c_L \neq c_T$. The case where the transverse and longitudinal speeds of sound are equal is simpler, since then we don't need the projection operators. 

A key relation we will need to show this cancellation comes from the symmetry transformation in \Eq{delta_pi}.  Since this is an invariance of the Lagrangian, this symmetry transformation relates the inverse propagator and three-point vertex, \eq{
\lim_{q \to 0}\alpha_i[V^{i\,  i_a}_{3\,\,\,\,\,\,\,j_a}(q,p_a)]+\beta^{i_a}_{\,\,\,\,\,k_a}[\Delta^{-1}(p_a)^{k_a}_{\,\,\,\,\,j_a}]=0 \, ,
}{V3_delta_relation}
which is even satisfied off-shell. In the following, we will assume that $\alpha$ and $\beta$ are linear operators, which is indeed the case for all the examples we will discuss in this paper.

There are two types of off-shell contributions that will vanish on-shell.  The first is the dispersion relation for a specific external particle $a$.  Consider corrections to the amplitude of the form
\eq{
\delta A^{i_1 \cdots i_n }_n =(\omega_a^2-c^2_a p_a^2)\Pi^{a}(p_a)^ {i_a}_{\,\,\,\,\,j_a} \mathcal O^{i_1 \cdots j_a \cdots i_n }_{n} \, ,
}{}
where we have inserted a projector $\Pi^a$ that enforces that leg $a$ has the correct corresponding longitudinal or transverse polarization, thus making the on-shell condition manifest.  By construction, $\delta A^{i_1 \cdots i_n }_n$ vanishes on-shell.  Next, we apply the right-hand side of the soft theorem in \Eq{soft_theorem_exact} to the above expression and then apply the on-shell conditions, yielding
\eq{
& -\Pi^{a}(p_a)^{i_a}_{\,\,\,\,\,j_a}\Big[\lim_{q \to 0} \alpha_i \left(  V^{i\, j_a\,k_a}_{3}\right) + \beta^{j_a\, k_a}   \left( \omega_a^2-c^2_{a} p_a^2\right)\Big]  \Pi^{a}(p_a)_{k_a \,l_a} \mathcal O^{i_1 \cdots l_a \cdots i_n }_{n} \,  ,
}{offshell_terms1}
at the relevant order in the soft expansion. To show that \Eq{offshell_terms1} vanishes, we sandwich \Eq{V3_delta_relation} between two projectors for particle $a$. Suppressing the indices, we obtain the relation
\eq{
\Pi^{a}\left[\lim_{q \to 0}\alpha( V_3 )\right]\Pi^{a}=-&\Pi^{a} [\beta( \Delta^{-1} )]\Pi^{a}\\
=-&\Pi^{a} \big[ \beta \left((\omega^2-c_a^2 p^2) \Pi^a + (\omega^2-c_{\bar a}^2 p^2) \Pi^{\bar a} \right) \big]\Pi^{a}\\
=-&\Pi^{a} [\beta(\omega^2-c_{a} ^2 p^2) ] \Pi^{a} \, ,
}{alphaV3_first}
where $\bar a$ denotes the mode orthogonal to $a$. To obtain the last line in \Eq{alphaV3_first}, we have used that $\Pi^{ a} \Pi^{\bar a} = 0$ and $\Pi^{a} [ \beta (\Pi^{a})] \Pi^{a}=\Pi^{a} [ \beta (\Pi^{\bar a})] \Pi^{a}=0$. This can be seen from
\eq{
\Pi^a[ \beta( \Pi^b)]\Pi^a &=\Pi^a[\beta (\Pi^{b 2})]\Pi^a
=\Pi^a\big[ \Pi^b\beta( \Pi^b)+\beta( \Pi^b) \Pi^b\big]\Pi^a  \, ,
}{eq: boost projectors}
which vanishes both when $b=a$ and $b=\bar a$.
Hence the sum of field basis dependent contributions in \Eq{offshell_terms1} vanishes.

The second off-shell contribution we consider, when the speeds of the two types of modes are different, $c_L \neq c_T$, is
\eq{
\delta A^{i_1 \cdots i_n }_n = \Pi^{\bar a}(p_a)^ {i_a}_{\,\,\,\,\,j_a} \mathcal O^{i_1 \cdots j_a \cdots i_n }_{n} \, ,
}{}
corresponding to contributions that vanish due to the projectors coming from the choice of external polarizations. As before, $\bar a$ denotes the mode orthogonal to $a$, so $\delta A^{i_1 \cdots i_n }_n$ again vanishes on-shell. First applying the soft theorem followed by the on-shell condition, we obtain
\eq{
&-\Pi^{a}(p_a)^{i_a}_{\,\,\,\,\,j_a}\left[\lim_{q \to 0} \alpha_i\left( V^{i\, j_a\,k_a}_{3}\right) \Delta^{\bar a }(p_a)_{k_a\,l_a} + \beta^{j_a\, k_a}   \left( \Pi^{\bar a}(p_a)_{k_a \,l_a} \right) \right] \mathcal O^{i_1 \cdots l_a \cdots i_n }_{n}  \, .
}{offshell_terms2}
To show that these terms vanish, we sandwich \Eq{V3_delta_relation} between $\Pi^a$ and $\Delta^{\bar a}$, which yields
\eq{
\Pi^a\left[\lim_{q \to 0} \alpha( V_3 )\right]\Delta^{\bar a} &= -\Pi^a[\beta( \Delta^{-1} )]\Delta^{\bar a} \\
&=-\Pi^a\Big[\beta \left( (\omega^2-c_a^2 p^2) \Pi^a + (\omega^2-c_{\bar a}^2 p^2) \Pi^{\bar a} \right) \Big]\Delta^{\bar a}\\
&=-\Pi^a\Big[  (\omega^2-c_a^2 p^2) \beta(\Pi^a)+(\omega^2-c_{\bar a}^2 p^2) \beta(\Pi^{\bar a})  \Big]\Delta^{\bar a}  \, ,
}{alphaV3_second}
where again we have used $\Pi^a \Pi^{\bar a} = 0$ to get to the third line in \Eq{alphaV3_second}. The first term in the square brackets in the third line in \Eq{alphaV3_second} vanishes on-shell.
Hence we obtain
\eq{
\Pi^a\left[ \lim_{q \to 0} \alpha( V_3) \right]\Delta^{\bar a}=-\Pi^a\beta[\Pi^{\bar a}]\Pi^{\bar a} \, ,
}{}
so that the off-shell terms in \Eq{offshell_terms2} cancel out. This shows that the soft theorem in \Eq{soft_theorem_exact} commutes with the on-shell conditions, thereby guaranteeing the field basis independence of the soft theorem.

\section{Superfluids}
\label{sec:Superfluids}

To start, we will consider the soft theorems for superfluids corresponding to nonlinearly realized time translations and Lorentz boosts.  These constrain the ${\cal O}(q^0)$ and ${\cal O}(q^1)$ terms in the amplitude in the soft limit, respectively.  Note that the latter was exhaustively studied in the interesting recent work of \cite{Green:2022slj} in the context of single field inflation, which in the flat space limit is described by the superfluid EFT \cite{Cheung:2007st}.  For completeness, we recapitulate results for superfluids here even though the important insights on this theory were discussed already in \cite{Green:2022slj}.

\subsection{Setup}

The superfluid EFT arises from spontaneous symmetry breaking an internal $U(1)$ symmetry where the phase degree of freedom $\phi$ has a time-dependent vacuum expectation value (VEV),\footnote{Equivalently, the VEV of the field transforming linearly under internal $U(1)$ is  $\langle e^{i\mu \phi} \rangle =e^{i\mu t}$, where $\mu$ is the chemical potential.}
\eq{
        \langle \phi \rangle =t \,.
}{}
The VEV spontaneously breaks $U(1)$ symmetry and time translations down to a diagonal subgroup. Lorentz symmetry is also spontaneously broken. The fluctuations around the VEV are described by a field $\pi$, so
\eq{
        \phi  = t + \pi \,.
}{eq:superfluid_phi}
Under time translations and Lorentz boosts, the field $\pi$ transforms nonlinearly,
\begin{align}
	\label{parameter_translation_superfluid}
         \text{time translations:} \quad   &\pi(x) \to \pi'(x')=\pi(x) + T \,, \\
	\text{Lorentz boosts:} \quad & \pi(x) \to \pi'(x')=\pi(x) + \vv v_i  \vv x^i + \vv v_i \left(\vv  x^i \partial_t  + t \vv \nabla^i  \right) \pi(x) \,,
	\label{parameter_boost_superfluid}
\end{align}
where $T$ and $\vv v$ are constant parameters.

As is well-known, the superfluid Lagrangian can be written in terms of the spacetime translation and Lorentz invariant combination,
\begin{align}
        X =-\frac12( \partial_\mu \phi\partial^\mu \phi +1 ) =  \dot\pi - \frac12 \partial_\mu \pi \partial^\mu \pi = \dot\pi + \frac12 \dot \pi^2 -  \frac12 (\partial_i \pi)^2 \, .
\end{align}
Considering terms with the fewest possible derivatives per field, we write down the leading terms in the superfluid EFT Lagrangian,
\eq{
        {\cal L}_{\rm superfluid}  &= M_1  X + \frac{M_2}{2} X^2 + \frac{M_3}{3!}X^3 + \cdots \\
                        & = M_1 \dot \pi + \frac{M_1+M_2}{2}  \dot \pi^2 -\frac{M_1}{2} (\partial_i \pi)^2  + \frac{M_3}{3!}  \dot \pi^3  + \frac{M_2}{2} \dot \pi (\dot \pi^2 - (\partial_i \pi)^2)  + \cdots \\
                        &= c^2 \dot \pi + \frac{1}{2} ( \dot \pi^2 -c^2 (\partial_i \pi)^2)  + \frac{g_3}{3!}  \dot \pi^3  + \frac{c^{-2}-1}{2} \dot \pi (\dot \pi^2 -c^2 (\partial_i \pi)^2) + \dots \,,
}{L_superfluid}
where $c$ is the speed of sound, $g$ is the coupling in the three-point on-shell amplitude, and the canonically normalized Lagrangian parameters are
\eq{
M_1 &= c^2 \,, \quad
M_2 &= 1-c^2 \,, \quad
M_3 &= g_3 +3 \frac{(1-c^2)^2}{c^2} \,.
}{}
As is common for spontaneously broken spacetime symmetries, the interactions are related to the speed of sound by symmetry.

The Feynman rules for the superfluid are trivial to derive from the Lagrangian in \Eq{L_superfluid}.
The propagator for the superfluid scalar is
\eq{
        \Delta(p) = \frac{1}{\omega^2 - c^2 \vv p^2}\,.
}{propagator_superfluid} 
Given the convention defined in \Eq{EOM}, the cubic interaction vertex is 
\eq{
	V_3 = ig_3 \omega_1 \omega_2 \omega_3 + i(c^{-2} -1) \left( \omega_1 (\omega_2 \omega_3 - c^2  \vv p_2 \cdot \vv p_3 )  +\textrm{cyclic} \right)\,.
}{3pt_superfluid}

\subsection{Amplitudes}
\label{sec:SuperfluidsAmp}

For completeness, let us summarize here some amplitudes describing the scattering of superfluid modes.  For example, the three-point scattering amplitude is
\eq{
        A_3 &= i  g_3  \omega_1 \omega_2 \omega_3 \,.
}{A3_superfluid}
For later convenience, let us define the kinematic variables
\eq{
        \omega_{ab}^2 &= (\omega_a + \omega_b)^2\,,\\
        s_{ab} &= (\omega_a + \omega_b)^2 - c^2 (\vv p_a + \vv p_b)^2\,,
}{eq:mandelstams}
where $s_{ab}$ reduces to the familiar Mandelstam variables for $c=1$. \
The four-point amplitude is
\begin{align}
        A_4 =&-g_3^2 \left(\frac{\omega_{12}^2}{s_{12}} + \frac{\omega_{13}^2}{s_{13}} + \frac{\omega_{14}^2}{s_{14}} \right)  \omega_1 \omega_2 \omega_3 \omega_4 +\frac{ 1-c^{2}}{4 c^4} (s_{12}^2 + s_{13}^2 + s_{14}^2)   \\
             & + \frac{g_3}{2c^2} (\omega_{12}^2s_{12}  +\omega_{13}^2s_{13} + \omega_{14}^2s_{14})  + g_4 \omega_1 \omega_2 \omega_3 \omega_4 \,, \nonumber
\end{align}
where $g_4=M_4 - 14g_3(1-c^2)/c^2 - 15(1-c^2)^3/c^4$. Note that the coefficient of the first term in $A_4$ is fixed by factorization. Naively, the coefficients of $(s_{12}^2 + s_{13}^2 + s_{14}^2)$ and $(\omega_{12}^2s_{12}  +\omega_{13}^2s_{13} + \omega_{14}^2s_{14})$ could have been independent contact terms. Nevertheless, the nonlinearly realized boost symmetry relates them to the speed of sound $c$ and the tree-point coupling $g_3$.

\subsection{Soft Theorem}

As noted earlier, the superfluid exhibits the spontaneous breaking of time translations as well as Lorentz boosts.  Let us now derive the soft theorems correspond to each of these broken symmetries.  To achieve this, we take the general form of the soft theorem in \Eq{soft_theorem_exact}, plug in the cubic interaction vertex $V_3$ for the superfluid, and then insert the $\alpha$ and $\beta$ parameters corresponding to either time translations or Lorentz boosts.  These will constrain the ${\cal O}(q^0)$ and ${\cal O}(q^1)$ terms in the amplitudes, respectively.

\subsubsection{Time Translations}
For the case of time translations, we see by inspection from \Eq{parameter_translation_superfluid} that  the symmetry transformation parameters $\alpha$ and $\beta$ defined in  \Eq{delta_pi} are simply
\begin{equation}
        \alpha = T \qquad \text{and} \qquad \beta = 0\,.
\end{equation}
Hence, the corresponding soft theorem is
\eq{
\lim_{q\rightarrow 0}  A_{n+1}(p_1, \cdots, p_n, q)  &= - \lim_{q\rightarrow 0} \sum_{a=1}^n   V_3(q,p_a) \Delta(p_a+q) A_{n}(p_1, \cdots, p_a+q, \cdots, p_n) \\
 & = \frac{1}{2} \sum_{a=1}^n \frac{i g  \omega \omega_a^2}{  \omega \omega_a -  c^2\, \vv q\cdot \vv p_a} A_n(p_1, \cdots, p_n)\,.
}{soft_theorem_superfluid_translation}
To derive the second line in \Eq{soft_theorem_superfluid_translation} we have used that
\eq{
\Delta(p_a+q) = \frac{1}{(\omega_a + \omega)^2 - c^2 (\vv p_a + \vv q)^2} = \frac{1}{2(\omega \omega_a - c^2\vv q\cdot \vv p_a)}  \,,
}{}
and that the three-point interaction vertex with two legs on-shell,  
\eq{
        V_3(q, p_a)   &\overset{ \hphantom{q \to 0}}{=}  -ig \omega \omega_a (\omega_a + \omega) +  2i (c^{-2} -1) (\omega_a + \omega) (\omega \omega_a - c^2 \vv q\cdot \vv p_a) \\
                                               &\overset{q \to 0}{=}   - ig \omega \omega_a^2 +  2i(c^{-2} -1)  \omega_a (\omega \omega_a - c^2 \vv q\cdot \vv p_a) \,,
}{}
where in the second line we have taken the soft limit.
 Plugging the above expression to \Eq{soft_theorem_superfluid_translation}, we see that
the first term persists while the total contribution from the second term, after cancellations with corresponding propagators, vanishes due to energy conservation $\sum_a \omega_a = 0$.

\subsubsection{Lorentz Boosts}

Next, we consider the soft theorem arising from the spontaneously broken Lorentz boosts. The parameters in \Eq{delta_pi} are identified by comparing with \Eq{parameter_boost_superfluid}, giving
\eq{
	\alpha = \vv v_i \vv x^i
	\qquad 
	\textrm{and}
	\qquad
	\beta = \vv v_i \left( \vv x^i \partial_t + t \vv \nabla^i \right) \,.
}{superfluid_boost_alpha_beta}
By specifying to the values in \Eq{superfluid_boost_alpha_beta} for the general soft theorem in \Eq{soft_theorem_exact}, we get 
\begin{align}
        \lim_{q\rightarrow0} \tfrac{\partial}{\partial \vv q^{i}} \left[ A_{n+1}( p_1,\cdots,p_n,q) \right] 
	=& - \lim_{q\rightarrow 0} \tfrac{\partial}{\partial \vv q^{i}} \sum_{a=1}^{n}  \left[ V_{3}(q,p_a) \Delta(p_a + q) A_{n}(p_1,\cdots ,p_a + q, \cdots , p_n) \right] \nonumber \\
	 & - \sum_{a=1}^{n} i \left(\omega_a \tfrac{\partial}{\partial \vv p_{ai}} + \vv p^i_a \tfrac{\partial}{\partial\omega_a} \right) \left[ A_{n}(p_1,\cdots,p_n) \right] \,,
\label{soft_theorem_superfluid_boost}
\end{align}
where the propagator and three-point vertex are those in eqs \eqref{propagator_superfluid} and \eqref{3pt_superfluid} respectively, and we have stripped the constant vector $v$.

Our derivation of this soft theorem only differs from that in \cite{Green:2022slj} in the field basis chosen, which changes the form of $V_3$ and $\beta$. The basis chosen in \cite{Green:2022slj} is related to ours by
\eq{
        \pi \to \pi - (c^{-2}-1) \pi \dot \pi\,.
}{}
In such basis the three-point vertex is
\eq{
	V^{\prime}_3 = i g \omega_1 \omega_2 \omega_3\,,
}{3pt_superfluid_otherbasis}
and the symmetry transformation is
\eq{
	\alpha^{\prime} = \vv v_i \vv x^i
	\qquad 
	\textrm{and}
	\qquad
	\beta^{\prime} = \vv v_i \left( c^{-2} \vv x^i \partial_t + t \vv \nabla^i \right) \,,
}{superfluid_boost_alpha_beta_otherbasis}
so the soft theorem is as derived in \cite{Green:2022slj}
\begin{align}
        \lim_{q\rightarrow0} \tfrac{\partial}{\partial \vv q^{i}} \left[ A_{n+1}( p_1,\cdots,p_n,q) \right] 
	=& - \lim_{q\rightarrow 0} \tfrac{\partial}{\partial \vv q^{i}} \sum_{a=1}^{n}  \left[ V^{\prime}_{3}(q,p_a) \Delta(p_a + q) A_{n}(p_1,\cdots ,p_a + q, \cdots , p_n) \right] \nonumber \\
	 & - \sum_{a=1}^{n} i \left(c^{-2}\omega_a \tfrac{\partial}{\partial \vv p_{ai}} + \vv p^i_a \tfrac{\partial}{\partial\omega_a} \right) \left[ A_{n}(p_1,\cdots,p_n) \right] \,.
\label{soft_theorem_superfluid_boost_otherbasis}
\end{align}
The above version of the soft theorem features a boost operator that is nonrelativistic, whereas the one in \Eq{soft_theorem_superfluid_boost} is relativistic. The former has the advantage that it annihilates the on-shell condition $\omega_a^2 - c^2 p_a^2 = 0$, thus making invariance under field redefinitions more manifest. Nevertheless, as explained in Sec.~\ref{sec:FR}, the soft theorem can be written in any basis.  

\section{Solids}
\label{sec:Solids}

\subsection{Setup}

The Lagrangian description for solids utilizes a three-vector field which acquires a vacuum expectation value,
\eq{
\langle \vv \phi^i\rangle = \vv x^i  \, ,
}{}
which spontaneously breaks part of the Poincar\'e symmetry. Fluctuations of this field are the NGBs for symmetry breaking, defined via
\eq{
\vv \phi^i  = \vv x^i + \vv \pi^i  \, ,
}{phonon_fluctuation_definition}
where $\pi$ is the phonon field.  Under spatial translations, the phonon transforms as 
\eq{
\textrm{spatial translation:}\quad & \vv \pi^i(x) \rightarrow \pi'^i(x')=\vv \pi^i(x) + \vv w^i \, ,
}{translation_phonon}
for a constant vector $\vv w$.  At the same time, the phonon transforms nonlinearly under boosts as
\eq{
\textrm{Lorentz boost:}\quad& \vv \pi^i(x) \rightarrow \pi'^i(x')= \vv \pi^i(x) + \vv v^i t +  \vv v_j  \left(\vv x^j \partial_t + t \vv \nabla^j \right) \vv \pi^i(x) \, ,
}{boost_phonon}
for a constant vector $\vv v$.
Here the first term on the right-hand side arises because $\pi$ transforms under boosts exactly like spatial position, as implied by \Eq{phonon_fluctuation_definition}.  The second term on the right-hand side arises because the spacetime argument of the phonon field actively transforms under boosts also.

To construct a Lagrangian that is invariant under nonlinearly realized boosts, we follow the procedure of \cite{Endlich:2012pz,Nicolis:2015sra} and define
\eq{
B^{ij} &= \partial_\mu \vv \phi^i \partial^\mu \vv \phi^j -\delta^{ij} = \vv \nabla^i \vv \pi^j +\vv \nabla^j \vv \pi^i + \partial_\mu \vv \pi^i \partial^\mu \vv \pi^j \, .
}{}
In three spatial dimensions, the only independent scalar components of this matrix are $[B]$, $[B^2]$, and $[B^3]$, where the square brackets denote a trace over spatial indices.  Hence, the general Lagrangian for the phonon is
\eq{
L_{\rm solid} &=  \sum_{i =0}^\infty  \sum_{j =0}^\infty  \sum_{k =0}^\infty \lambda_{ijk} [B]^{i} [B^2]^{j} [B^3]^{k}  \, ,
}{L_solid}
corresponding to the modes of a solid.

The three components of the phonon can be further decomposed into a single longitudinal mode and two transverse  modes via
\eq{
        \vv \pi^i = \vv \pi^i_L + \vv \pi^i_T \qquad \textrm{where} \qquad   \vv \nabla_i  \vv \pi^i_T=0  \qquad \textrm{ and } \qquad   \epsilon_{ijk} \vv \nabla^j   \vv \pi_L^k =0  \, .
}{LT_decomp}
By expanding the Lagrangian to quadratic order, we learn that the longitudinal and transverse speeds of sound, $c_L$ and $c_T$, are related to the coupling constants via
\eq{
\lambda_{010} = \tfrac14 (1-c_T^2) \qquad \textrm{and} \qquad
\lambda_{200} = -\tfrac18 (1+c_L^2-2c_T^2)  \, .
}{}
Otherwise, the couplings are completely unfixed by the spontaneous symmetry breaking pattern.\footnote{In general, there will be further thermodynamic constraints on the couplings, depending on the physical system of interest.}

The soft theorem in \Eq{soft_theorem_exact} depends on the propagator and cubic vertex, $\Delta$ and $V_3$.   Let us briefly present expressions for these quantities in  the case of a solid. The propagator for the phonons of the solid is given in \Eq{propagator}. 
From the solid Lagrangian in \Eq{L_solid}, we also compute the cubic interaction vertex, 
\eq{
V_3^{i_1 i_2 i_3} (p_1, p_2, p_3)=&-\tfrac{i}{2} (1+c_L^2-2c_T^2) \big((\omega_2 \omega_3- \vv p_2 \cdot\vv p_3 ) \vv p_1^{i_1}\delta^{i_2 i_3} \big)\\
& + i(1-c_T^2)\big((\omega_2 \omega_3- \vv p_2 \cdot \vv p_3 )( \vv p_1^{i_3}\delta^{i_1 i_2}+\vv p_1^{i_2}\delta^{i_1 i_3}) \big)\\
&-2i  \lambda_{001} \big(\vv p_1^{i_2}\vv p_2^{i_3} \vv p_3^{i_1}+3 (\vv p_2 \cdot \vv p_3 )p_1^{i_2} \delta^{i_1 i_3}) \big)\\
&-4i \lambda_{110}\big(\vv p_1^{i_1}\vv p_2^{i_3} \vv p_3^{i_2}+ (\vv p_2 \cdot \vv p_3 )p_1^{i_1}\delta^{i_2 i_3}) \big)\\
&-8i \lambda_{300}\vv p_1^{i_1}\vv p_2^{i_2}\vv p_3^{i_3} +\textrm{permutations} \, ,
}{V3}
which we can freely rewrite on the support of total momentum conservation.

\subsection{Amplitudes}

Next, let us briefly describe some explicit phonon amplitudes. The three-point scattering amplitudes for various combinations of longitudinal and transverse polarizations are
\eq{
A_{LLL} &= -i\frac{3}{c_L^3}((1- c_L^2)^2+16 (\lambda_{001}+\lambda_{110}+\lambda_{300})\big)\omega_1\omega_2\omega_3 \,, \\
A_{TTT} &= -i\frac{1}{c_T^2}\big((1- c_T^2)^2+6 \lambda_{001}\big) \omega_3 (\omega_1-\omega_2)(\vv e_1 \cdot \vv e_2)(\vv p_2  \cdot \vv e_3) + \text{cyclic} \,,\\
A_{LLT}&= i\frac{\omega_1^2-\omega_2^2}{2c_L^2 c_T^2 \omega_1 \omega_2 }(\vv p_1  \cdot \vv e_3)\bigg(2c_L^2c_T^2\big((c_T^2-c_L^2)(\omega_1^2+\omega_2^2)+\omega_1\omega_2(c_T^2-3c_L^2)\big)\\
& \hspace*{0.5in} +\big(2-c_T^2-3c_L^2+8(2\lambda_{110}+3\lambda_{001})\big) \big((c_T^2-c_L^2)(\omega_1^2 + \omega_2^2)-2c_L^2\omega_1\omega_2)\big)\bigg) \,,\\
A_{TTL}&= -i\frac{(\vv p_2 \cdot \vv e_1)( \vv p_1 \cdot \vv e_2)}{c_L c_T^2\omega_3}\bigg(2c_T^2(1-c_T^2+4\lambda_{110}+9\lambda_{001}) \omega_3^2\\
& \hspace*{0.5in}-c_L^2\big((1-c_T^4+6\lambda_{001})(\omega_1^2+\omega_2^2)+4c_T^2(1-c_T^2)\omega_1\omega_2\big)\bigg)\\
& \hspace*{0.1in}+i \frac{\vv e_1 \cdot \vv e_2}{2c_L^3 c_T^4}\omega_3\bigg(c_L^2c_T^2\big((c_T^2-c_L^2)(\omega_1^2+\omega_2^2)+2c_T^2(1-c_L^2+c_T^2)\omega_1\omega_2\big)\\
& \hspace*{0.5in} +c_L^2\big((1-c_T^2)^2+6 \lambda_{001}\big)\big((c_L^2-c_T^2)(\omega_1^2+\omega_2^2)-2c_L^2\omega_1\omega_2\big)\\
&\hspace*{0.5in} +c_T^2\big(c_T^2+8\lambda_{110}+6 \lambda_{001}\big)\big(c_L^2(\omega_1^2+\omega_2^2)-c_T^2\omega_3^2\big)\bigg) \,,
}{A3_solid}
where we have eliminated $\vv p_i \cdot \vv p_j$ using the minimal on-shell kinematic basis defined in \Eq{min_kin_3}.  As noted in \cite{Brauner:2022ymm}, for real on-shell kinematics the external three-momenta are necessarily collinear, so the three-point amplitudes with an odd number of transverse polarizations are zero. However, collinearity is avoided in the case of complex kinematics.

The four-point amplitude is immensely complicated so we do not write it explicitly.  However, it can be found in an ancillary {\tt Mathematica} notebook included with this paper containing all of our amplitudes.  

\subsection{Soft Theorem}

Next, to evaluate \Eq{soft_theorem_exact} we must specify $\alpha$ and $\beta$ which depend on which symmetry is spontaneously broken.  In what follows, we consider the case of spatial translations and Lorentz boosts, respectively.

\subsubsection{Spatial Translations}

By inspection, we see that the nonlinearly realized spatial translation in \Eq{boost_phonon} corresponds to  \Eq{delta_pi} by identifying 
\eq{
\alpha^i = w^i  \qquad \textrm{and} \qquad
\beta^i_{\,\,j} = 0 \, .
}{alpha_beta_solid_translation}
Plugging this into \Eq{soft_theorem_exact}, we obtain the soft theorem corresponding to spatial translations of a phonon in a solid,
\eq{
\lim_{q\rightarrow 0}  A^{i_1 \cdots i_n i}_{n+1}(p_1, \cdots, p_n, q)&=
 -\lim_{q\rightarrow 0} \sum_{a=1}^n    V^{i\,  i_a}_{3\,\,\,\,\,\,\,j_a}(q,p_a) \Delta^{j_a}_{\,\,\, \,\,k_a}(p_a+q) A^{i_1 \cdots j_a \cdots i_n }_{n}(p_1, \cdots, p_a+q, \cdots, p_n) \, ,
}{soft_theorem_solid_translation}
where we have stripped off the constant translation vector $w_i$, leaving a free $i$ index.

We have explicitly evaluated \Eq{soft_theorem_solid_translation} for the case of four- and three-point amplitudes and verified its validity. Here $A_3$ and $A_4$ should be evaluated in the minimal kinematic bases defined in \Eq{min_kin_3} and \Eq{min_kin_4}. Also, to evaluate the above expression one must, in the end, contract the polarization indices of the hard particles, $i_1, i_2, i_3$, with explicit polarizations which are either longitudinal or transverse.   The on-shell conditions for those legs should also correlate with the choice of polarizations, since the longitudinal and transverse speeds of sound are in general different.  By computing all possible combinations of longitudinal and transverse combinations for the external legs, we have verified that the above formula holds. 

Note that in general for the soft limit of $A_n$ with $n>4$ we do not encounter the fractional soft limits of \cite{Brauner:2022ymm}. In that setup, the authors assume real kinematics, for which taking the soft limit from four- to three-point yields collinear momenta for the latter. For five- and higher-point amplitudes the soft limit does not yield collinear kinematics in the amplitudes on the right-hand side of the soft theorem. Moreover, as noted earlier, we assume complex kinematics throughout, as is common in the study of gauge theory amplitudes.

\subsubsection{Lorentz Boosts}

Next, we consider the soft theorem corresponding to spontaneously broken Lorentz transformations.
Comparing the nonlinearly realized Lorentz boost of the phonon in \Eq{boost_phonon} to \Eq{delta_pi}, we see that the transformation parameters are
\eq{
\alpha^i = v^i t \qquad \textrm{and} \qquad
\beta^i_{\,\,j} = v_k \left(x^k \partial_t + t \nabla^k\right)\delta^i_{\,\,\,j} \, .
}{alpha_beta_solid_boost}
Inserting \Eq{alpha_beta_solid_boost} into \Eq{soft_theorem_exact}, we obtain the soft theorem corresponding to Lorentz boosts of a phonon in a solid,
\eq{
&\lim_{q\rightarrow 0}  \tfrac{\partial}{\partial\omega}  \left[ A^{i_1 \cdots i_n i}_{n+1}(p_1, \cdots, p_n, q) \right] =\\
 &\qquad\qquad -\lim_{q\rightarrow 0} \sum_{a=1}^n   \tfrac{\partial}{\partial\omega} \left[  V^{i\,  i_a}_{3\,\,\,\,\,\,\,j_a}(q,p_a) \Delta^{j_a}_{\,\,\, \,\,k_a}(p_a+q) A^{i_1 \cdots j_a \cdots i_n }_{n}(p_1, \cdots, p_a+q, \cdots, p_n)\right]\\
&\qquad\qquad -\sum_{a=1}^n i\left(\omega_a \tfrac{\partial}{\partial p_{ai}}+ p_a^i \tfrac{\partial}{\partial\omega_a}\right)  \left[ A^{i_1 \cdots i_a \cdots i_n }_{n} (p_1, \cdots, p_a, \cdots, p_n)\right] \, ,
}{soft_theorem_solid_Lorentz}
where the
propagator $\Delta$ and cubic vertex $V_3$ are defined in \Eq{propagator} and \Eq{V3}, respectively.  As before, we have stripped off the constant Lorentz boost vector $v_i$, leaving a free $i$ index.

We have verified by explicit calculation that the soft theorem relating the four-point and three-point amplitudes is satisfied. As before, in order to verify this soft theorem it is important to go to the minimal kinematic bases for $A_3$ and $A_4$ in \Eq{min_kin_3} and \Eq{min_kin_4}. Furthermore, we must contract with explicit longitudinal or transverse polarizations for the hard external states.  Performing this for all combinations, we find that the above soft theorem is indeed satisfied.

\section{Fluids}
\label{sec:Fluids}

\subsection{Setup}

As emphasized in \cite{Nicolis:2015sra}, fluids are nothing more than solids with an enhanced symmetry. In particular, the Lagrangian for fluids is the same as the one for solids except with couplings constrained to exhibit an additional invariance under infinitesimal volume-preserving diffeomorphisms,
\eq{
\textrm{diffeomorphism:}\quad \phi^i \rightarrow \phi'^i=\phi^i + \xi^i(\phi) \, .
}{}
The volume-preserving condition implies that $\partial_i \xi^i=0$.
In terms of the physical  phonon field this corresponds to
\eq{
\textrm{diffeomorphism:} \quad \pi^i \rightarrow \pi'^i=\pi^i + \xi^i(x+\pi) = \pi^i + \xi^i(x) + \partial_j \xi^i(x) \pi^j + \cdots  \, ,
}{diff}
expanded to linear order in the phonon field. 

Invariance under volume-preserving diffeomorphisms implies that the fluid Lagrangian can only depend on the combination
\eq{
\det B' =\det (1+B)=1+[B]+\frac{1}{2}\left([B]^2-[B^2]\right)+ \frac{1}{3!} \left([B]^3 - 3 [B][B^2]+ 2 [B^3]\right)\,,
}{}
and thus it takes the form
\eq{
{\cal L}_{\rm fluid} = -\frac12 \text{det}B' + \tau_0\, \text{det}B'^2 + \tau_1\, \text{det}B'^3 + \tau_2\, \text{det}B'^4 + \cdots \, ,
}{}
where $\tau_0 = (1-c_L^2)/8$.
This implies that fluid dynamics are obtained by imposing the following constraints on the solid Lagrangian,
\eq{
\lambda_{010} &= \tfrac14 \,, \\
\lambda_{001} &= -\tfrac16 \,, \\
\lambda_{200} &= -\tfrac{1}{8}  (1 + c_L^2) \,,\\
\lambda_{020} &= \tfrac{1}{32} (1 - c_L^2) \,,\\
\lambda_{101} &= \tfrac{1}{12} (1 - c_L^2) \,,\\
\lambda_{110} &= \tfrac{1}{8}  (1 + c_L^2) \,,\\
\lambda_{210} &= -\tfrac{3}{16}(1 - c_L^2) -\tfrac{3}{2} \tau_1 \,,\\
\lambda_{300} &= \tfrac{1}{24} (1 - 3 c_L^2)+ \tau_1 \,,\\
\lambda_{400} &= \tfrac{7}{96} (1 - c_L^2) + \tfrac{3}{2} \tau_1 + \tau_2  \, ,
}{lambda_fluid}
where $\tau_1$ and $\tau_2$ are residual free parameters of the fluid Lagrangian.

As emphasized in \cite{Endlich:2010hf, Nicolis:2015sra}, the fluid EFT is peculiar because the transverse speed of sound is vanishing, so $c_T=0$.  Consequently, the transverse modes lack a gradient kinetic term and the corresponding degrees of freedom are not localized particles in any conventional sense.\footnote{We will not shed any new insight on this particular problem, though there has been recent progress making sense of perfect fluids in two dimensions by recasting volume-preserving diffeomorphisms as $SU(N)$ transformations as $N\rightarrow \infty$ \cite{Dersy:2022kjd}. Curiously, a similar construction yields a nonperturbative formulation double copy relating scalar EFTs in two dimensions \cite{Cheung:2022mix}.  This double copy structure also arises in certain non-Abelian generalizations of the Navier-Stokes equations \cite{Cheung:2020djz}.}  Instead, we will consider the fluid case as a mathematically well-defined limit of small $c_T \rightarrow 0$.  While strict vanishing may be ill-defined, the limit is a straightforward way to regulate the corresponding solid amplitudes on the approach to fluid dynamics.

\subsection{Amplitudes}

Next, we record explicit fluid phonon amplitudes. The three-point scattering amplitude of longitudinal modes is
\eq{
A_{LLL} &= -i\frac{\tilde\tau_1}{c_L^3}\omega_1\omega_2\omega_3 \,,
}{A3_fluid}
where $\tilde\tau_1 = 3((1- c_L^2)^2+16 \tau_1)$, whereas the four-point amplitude is
\eq{
A_{LLLL} &= -\frac{\tilde\tau_1^2}{c_L^6}
  \left(\frac{\omega_{12}^2}{s_{12}} + \frac{\omega_{13}^2}{s_{13}} + \frac{\omega_{14}^2}{s_{14}} \right)  \omega_1 \omega_2 \omega_3 \omega_4 + \frac{ 1-c_L^{2}}{4 c_L^2} (s_{12}^2 + s_{13}^2 + s_{14}^2)   \\
             & +\frac{\tilde\tau_1}{2c_L^4} (\omega_{12}^2s_{12}  +\omega_{13}^2s_{13} + \omega_{14}^2s_{14})  + \frac{\tilde\tau_2}{c_L^6} \omega_1 \omega_2 \omega_3 \omega_4 \,,
}{A4_fluid}
where $\tilde \tau_2 = 3 c_L^2\big(128 \tau_2 -5 (1 - c_L^2)^3 \big) + \tilde\tau_1 \big(3 \tilde\tau_1 + 10 c_L^2 (1 - c_L^2) \big) $.

\subsection{Soft Theorem}

\subsubsection{Diffeomorphisms}

In order to verify the soft theorem in \Eq{soft_theorem_exact}, we must compute the transformation parameters $\alpha$ and $\beta$ for volume-preserving diffeomorphisms, and the propagator and cubic vertex $\Delta$ and $V_3$ for the fluid.

To begin, let us series expand a general volume-preserving diffeomorphism in powers of the space coordinate,
\eq{
 \xi^i(x)  &= \sum_{a=1}^\infty \xi^i_{\,\,j_1 \cdots j_a} x^{j_1} \cdots x^{j_a} \qquad \textrm{where} \qquad \xi^i_{\,\, j_1 \cdots i \cdots j_a}=0  \, .
}{}
Here we will be interested in verifying the leading nontrivial component of the diffeomorphism, which is linear in the space coordinate
\eq{
 \xi^i(x)  &=  \xi^i_{\;\;j} x^j \qquad \textrm{where} \qquad \xi^i_{\,\,i}=0 \, .
}{}
Recasting this leading diffeomorphism in terms of the parameters 
in \Eq{delta_pi}, we find that
\eq{
\alpha^i =  \xi^i_{\;\;j} x^j \qquad \textrm{and} \qquad
\beta^i_{\,\,j} =  \xi^i_{\,\,j} \, .
}{}
Next, to obtain $\Delta$ and $V_3$ we simply take the expressions in \Eq{propagator} and \Eq{V3} for the solid and insert \Eq{lambda_fluid}.

Putting this all together, we learn that the soft theorem in \Eq{soft_theorem_exact} applied to the leading volume-preserving diffeomorphisms of a fluid is
\eq{
&\lim_{q\rightarrow 0}  \tfrac{\partial}{\partial q_j}  \left[ A^{i_1 \cdots i_n i}_{n+1}(p_1, \cdots, p_n, q) \right] =\\
 &\qquad\qquad -\lim_{q\rightarrow 0} \sum_{a=1}^n    \tfrac{\partial}{\partial q_j} \left[  V^{i\,  i_a}_{3\,\,\,\,\,\,\,j_a}(q,p_a) \Delta^{j_a}_{\,\,\, \,\,k_a}(p_a+q) A^{i_1 \cdots j_a \cdots i_n }_{n}(p_1, \cdots, p_a+q, \cdots, p_n)\right]\\
&\qquad\qquad -\sum_{a=1}^n  i \delta^{i i_a} \delta_{j_a}^j  \left[ A^{i_1 \cdots j_a \cdots i_n }_{n} (p_1, \cdots, p_a, \cdots, p_n)\right] + \textrm{terms proportional to } \delta^{ij}  \, ,
}{soft_theorem_fluid_diff}
where we have stripped off the constant diffeomorphism parameter $\xi_{ij}$, leaving free $i$ and $j$ indices.
Thus we see that the terms proportional to $\delta^{ij}$ are projected out when the left- and right-hand sides are contracted into $\xi_{ij}$, which is by construction traceless.

To verify the above soft theorem we compute the three- and four-point amplitudes for fluid phonons by imposing the conditions on coupling constants in \Eq{lambda_fluid} on our amplitudes for solid phonons. By explicit computation we have verified the validity of the above soft theorem relating the four- and three-point amplitudes. As before, this check requires going to minimal kinematic basis for $A_3$ and $A_4$.  Furthermore, to avoid pathologies involving transverse polarizations of external states, we restrict to the case where all external polarizations are longitudinal.

Note that in principle, one can also derive soft theorems for high-order diffeomorphisms.  In particular, we could consider the next-to-leading diffeomorphism defined by
\eq{
\alpha^i = \tfrac12 \xi^i_{\;\;jk} x^jx^k \qquad \textrm{and} \qquad
\beta^i_{\,\,j} =  \xi^i_{\,\,jk} x^k  \, ,
}{NLOdiffparam}
where $\xi^i_{\;\;ik}= \xi^i_{\;\;ki}=0$.  In this case, the corresponding soft theorem will involve the action of the differential operator, $\alpha^i= -\tfrac12 \xi^i_{\;\;jk} \tfrac{\partial}{\partial q_j}  \tfrac{\partial}{\partial q_k} $.  This effectively extracts ${\cal O}(q^2)$ terms from the amplitude.  
The general soft theorem in \Eq{soft_theorem_exact} applies for {\it any} spontaneously broken symmetry,  including next-to-leading diffeomorphism. We do not explicitly construct and evaluate the soft theorem for next-to-leading diffeomorphism in this paper, but it is straightforward to do so by inserting \Eq{NLOdiffparam} into \Eq{soft_theorem_exact}.

\section{Framids}
\label{sec:Framids}

\subsection{Setup}

The framid theory exhibits a minimal field content needed to represent the spontaneous breaking of Lorentz symmetry.  The setup centers on a four-vector field whose vacuum expectation value,
\eq{
\langle A_\mu (x)\rangle  = \delta_\mu^0 \, ,
}{}
spontaneously breaks Lorentz symmetry.  Fluctuations about this value are parameterized by framon fields which are the NGBs of boosts,
\eq{
A_\mu = \exp(i \vv \pi^j \vv K_j)_\mu^{\,\,\,\,\nu} \delta_\nu^0 \, ,
}{}
where $K^i$ is a three-vector parameterizing Lorentz boosts. Expanding in powers of the fields, we obtain explicit formulas for the four-vector field,
\eq{
A_0 &= 1 + \tfrac12 \vv \pi^2 +\cdots \,, \\
\vv A_i &= \vv \pi_i(1+ \tfrac16  \vv \pi^2  +\cdots) \,.
}{}
By construction, boosts are nonlinearly realized as constant shifts of the framon field,
\eq{
\textrm{Lorentz boost:}\quad& \vv \pi^i(x) \rightarrow \vv \pi'(x')=\pi^i(x) + \vv v^i  +  \vv v_j \left(\vv x^j \partial_t + t \vv \nabla^j \right) \vv \pi^i(x) \, ,
}{boost_framid}
where, as before, the last term on the right-hand side appears because the spacetime position of the framon is boosted in the transformation.  Note that the framon does not nonlinearly realize translation symmetries.

The leading order boost invariant Lagrangian for the framon is
\eq{
{\cal L}_{\rm framid} =  -\tfrac12
M_3^2 (\partial_\mu A^\mu)^2  -\tfrac12 M_2^2 (\partial_\mu A_\nu)^2 -\tfrac12 (M_2^2 - M_1^2)(A^\rho \partial_\rho A_\mu)^2  \, .
}{}
As before, we can expand to quadratic order in the framons in order to express some of the couplings in terms of the speeds of sound of the longitudinal and transverse modes,
\eq{
c_T^2 = \frac{M_2^2}{M_1^2} \qquad \textrm{and} \qquad
c_L^2 = \frac{M_2^2 + M_3^2}{M_1^2}  \, .
}{}

In order to evaluate the soft theorem in \Eq{soft_theorem_exact} we must compute the propagator $\Delta$ and cubic vertex $V_3$ of the framid theory.  Conveniently, the framon has an identical dispersion relation to the phonon, so $\Delta$ is defined as in \Eq{propagator}. 

 Meanwhile, the cubic interaction vertex is straightforwardly extracted from ${\cal L}_{\rm framid}$, yielding 
\eq{
V_3^{i_1 i_2 i_3} (p_1, p_2, p_3) =& - (1-c_T^2)\left(\omega_1 (\delta^{i_1 i_2} \vv p_2^{i_3}+\delta^{i_1 i_3} \vv p_3^{i_2}) \right)\\
&- (c_T^2-c_L^2)\left(\omega_1 (\delta^{i_1 i_2} \vv p_3^{i_3}+\delta^{i_1 i_3} \vv p_2^{i_2})\right) + \textrm{cyclic} \, .
}{V3_framid}
As noted in \cite{Nicolis:2015sra}, in the relativistic limit of $c_L=c_T=1$, the framid coincides with the nonlinear sigma model (NLSM), which is why $V_3$ vanishes in this case. 

\subsection{Amplitudes}

The three-point on-shell scattering amplitudes for framons are
\eq{
A_{LLL} &= \frac{(1- c_L^2)}{c_L}(\omega_1^2+\omega_2^2+\omega_3^2) \,, \\
A_{TTT} &= (1-c_T^2)(\omega_1-\omega_2)( \vv e_1 \cdot \vv e_2)( \vv p_1 \cdot \vv e_3) + \text{cyclic} \,, \\
A_{LLT}&= \omega_1\left(\tfrac{(c_L^2-c_T^2)}{c_T^2} \tfrac{\omega_3^2}{\omega_1\omega_2}    +(1-c_L^2)   \right) ( \vv p_1 \cdot \vv e_3) + (1\leftrightarrow 2) \,,\\
A_{TTL}&= 3c_L(1-c_T^2)( \vv p_2 \cdot \vv e_1)( \vv p_1 \cdot \vv e_2)+\big(\tfrac{(c_L^2-c_T^2)(1-3c_T^2)}{2c_L c_T^2}\omega_3^2-\tfrac{2c_L(1-c_T^2)}{ c_T^2}\omega_1\omega_2\big)(\vv e_1 \cdot \vv e_2) \,,
}{A3_framid}
while our four-point amplitude can be found in the attached ancillary {\tt Mathematica} file.

\subsection{Soft Theorem}

\subsubsection{Lorentz Boosts}

Comparing \Eq{delta_pi} to \Eq{boost_framid}, we see that the transformation parameters corresponding to the nonlinearly realized Lorentz boosts of the framon are
\eq{
\vv \alpha^i = \vv v^i  \qquad \textrm{and} \qquad
\beta^i_{\,\,j} = \vv v_k \left( \vv x^k \partial_t + t \vv \nabla^k \right) \delta^i_{\,\,\,j}  \, .
}{alpha_beta_framid}
Combining this with \Eq{soft_theorem_exact}, we obtain 
\eq{
&\lim_{q\rightarrow 0}   A^{i_1 \cdots i_n i}_{n+1}(p_1, \cdots, p_n, q)=\\
 &\qquad\qquad -\sum_{a=1}^n     V^{i\,  i_a}_{3\,\,\,\,\,\,\,j_a}(q,p_a) \Delta^{j_a}_{\,\,\, \,\,p_a}(p_a+q) A^{i_1 \cdots j_a \cdots i_n }_{n}(p_1, \cdots, p_a+q, \cdots, p_n)\\
&\qquad\qquad -\sum_{a=1}^n \left(\omega_a \tfrac{\partial}{\partial \vv p_{ai}}+ \vv p_a^i \tfrac{\partial}{\partial\omega_a}\right)  \left[ A^{i_1 \cdots i_a \cdots i_n }_{n} (p_1, \cdots, p_a, \cdots, p_n)\right]  \, ,
}{soft_thm_framid}
which is the soft theorem corresponding to boosts in the framid.

By explicit calculation, we have verified the framon soft theorem at five-, four- and three-point, by using a minimal kinematic basis and plugging in all possible combinations of longitudinal or transverse polarizations for the hard external legs.

\section{Soft Bootstrap}
\label{sec:bootstrap}

Our analysis thus far has focused on incarnations of the soft theorem in \Eq{soft_theorem_exact}, which relate {\it nonzero} expressions involving the  $(n+1)$-point and $n$-point amplitudes.  However, in the special circumstance where soft limits vanish---also known as Adler zeros---the corresponding theories typically exhibit enhanced symmetry structures.  
Concretely, the Adler zero stipulates that
\eq{
\lim_{q\rightarrow 0}   A_{n+1}(p_1,\cdots ,p_n,q) = \mathcal{O}(q^1) \,,
}{Adler}
which is the case for, e.g., amplitudes of pions in the NLSM \cite{Adler:1964um}. 
In special circumstances, scalar EFTs can exhibit an {\it enhanced} Adler zero,
\eq{
\lim_{q\rightarrow 0}   A_{n+1}(p_1,\cdots ,p_n,q) = \mathcal{O}(q^2) \,.
}{enhanced_soft}
This is the case for Dirac-Born-Infeld (DBI) theory and the Galileon. Remarkably, the NLSM, DBI, and the Galileon exhibit a soft behavior of amplitudes that is enhanced beyond what is naively expected simply from counting the number of derivatives per interaction vertex. 
Hence, by writing an ansatz and imposing \Eq{Adler} or \Eq{enhanced_soft} as constraints, one can bootstrap these theories from first principles \cite{Susskind:1970gf,Cheung:2014dqa, Cheung:2016drk,Cheung:2015ota}.
These resulting theories have highly constrained interactions, and were dubbed {\it exceptional} scalar EFTs.   The  soft bootstrap has also been extended to broader classes of theories, including theories with vectors or fermions \cite{Kampf:2013vha, Luo:2015tat,Elvang:2017mdq,Elvang:2018dco,Cheung:2018oki,Low:2014nga,Low:2015ogb,Low:2017mlh,Low:2018acv,Liu:2019rce,Low:2019ynd,Dai:2020cpk,Roest:2019oiw,Kampf:2021bet,Kampf:2021tbk} as well as nonrelativistic theories \cite{Mojahed:2022nrn,Pajer:2020wnj,Stefanyszyn:2020kay}.

In the context of spacetime symmetry breaking, it is natural to ask: Are there exceptional nonrelativistic EFTs? Are there exceptional theories of phonons and framons?

\subsection{Exceptional Phonons}

We start by considering effective theories of phonons, i.e., corresponding to spacetime symmetry breaking pattern of solids, fluids, and superfluids, which all have interaction vertices with one derivative per field. First, we want to establish the Adler zero for these theories, i.e., the requirement that the amplitudes vanish as ${\cal O}(q^1)$ in the soft limit.

For a theory with one derivative per field, all interaction vertices involving the soft particle will scale as ${\cal O}(q^1)$. This suggests that the on-shell amplitudes should also scale as ${\cal O}(q^1)$, thus exhibiting an Adler zero.  However, as is well-known, this reasoning fails in the presence of three-point interaction vertices, which generically induce ${\cal O}(q^{-1})$ soft poles. These contributions can, in principle, conspire with ${\cal O}(q^1)$ contributions from the interaction vertex to give an amplitude that scales as ${\cal O}(q^0)$.

In relativistic theories of derivatively coupled scalars, there are no such soft poles because there are no on-shell three-point amplitudes. The only nonzero three-point amplitude for relativistic scalars is a constant arising from a cubic potential term, which is absent by definition for derivatively coupled scalars. The absence of relativistic three-point scalar amplitudes implies the existence of a field basis where the corresponding three-point vertex is zero and thus there are no singular terms in the soft limit. 

In contrast, nonrelativistic theories can have nontrivial three-point amplitudes, even in theories with interaction vertices with one derivative per field. Thus, the Adler zero is not automatic. A sufficient condition for having an Adler zero for such nonrelativistic theories is to demand that all three-point amplitudes vanish.

\subsubsection{Solids}

Let us begin by imposing an Adler zero for phonons in a solid. The three-point solid amplitudes are given in \Eq{A3_solid}.
Demanding that all of these vanish for any choice of external modes, transverse or longitudinal, implies a universal speed of sound, \footnote{Note that this constraint is not compatible with thermodynamic constraints imposed by bulk stability in ordinary solids \cite{Landau:1986aog}.}
\eq{
        c_T = c_L = c\,,
}{eq:condsolids1}
in addition to the constraints
\eq{
 6 \lambda_{001} = - 8\lambda_{110} = 48 \lambda_{300} = -(1-c^2)^2\,.
}{eq:condsolids2}
Given the restrictions in Eqs.~\eqref{eq:condsolids1} and \eqref{eq:condsolids2}, the following field redefinition
\eq{
\pi^i \to \pi^i + (1-c^2) \pi^j \partial_j \pi^i  \, ,
}{}
sets the three-point vertex to zero. In this basis the soft theorem for spontaneously broken spatial translations in Eq.~\eqref{soft_theorem_solid_translation} with $V_3=0$ shows the existence an Adler zero for $n$-point amplitudes.

One might ask whether such restrictions on the solid couplings are technically natural. In other words, do the choices of couplings in Eqs.~\eqref{eq:condsolids1} and \eqref{eq:condsolids2} enhance the symmetries of the solid EFT? Unfortunately, the answer is no. While the vanishing of the three-point amplitude and a relativistic dispersion relation suggest a possible emergent Lorentz symmetry, this is trivially broken by higher-point amplitudes. 

Next, it is natural to further impose the condition that the ${\cal O}(q^1)$ term in the amplitude vanishes, yielding an enhanced ${\cal O}(q^2)$ soft limit. A necessary condition for this is given by the soft theorem for spontaneously broken boosts. In the basis where the three-point vertex is zero it takes the form
\eq{
&\lim_{q\rightarrow 0}  \tfrac{\partial}{\partial\omega}  \left[ A^{i_1 \cdots i_n i}_{n+1}(p_1, \cdots, p_n, q) \right] =-\sum_{a=1}^n i\left(\omega_a \tfrac{\partial}{\partial p_{ai}}+ c^2 p_a^i \tfrac{\partial}{\partial\omega_a}\right)  \left[ A^{i_1 \cdots i_a \cdots i_n }_{n} (p_1, \cdots, p_n)\right] \, .
}{eq:softboostNR}
Interestingly, the enhanced Adler zero requires choosing couplings in the solid EFT such that the theory has an emergent boost symmetry with respect to the speed of sound $c$.

The boost soft theorem in \Eq{soft_theorem_solid_Lorentz} does not capture all terms of ${\cal O}(q)$ in the soft expansion of the solid amplitude. Hence, the emergence of a relativistic symmetry is not sufficient to guarantee an enhanced Adler zero. 
By explicitly imposing the enhanced Adler zero on the four- and five-point scattering amplitudes of longitudinal phonons in a solid, we  constrain the couplings in the solid Lagrangian in \Eq{L_solid} according to
\eq{
252\lambda_{101} &= 672\lambda_{020} = 1152\lambda_{400} = -224\lambda_{210} = 21(1-c^2)^3\,,\\
192\lambda_{310} &=
320\lambda_{120} =
-240\lambda_{201} =
-120\lambda_{011} =
-768\lambda_{500} =5(1-c^2)^4\,.
}{4pt5pt_bootstrap}
These conditions must be imposed in addition to the constraints in Eqs.~\eqref{eq:condsolids1} and \eqref{eq:condsolids2} which are needed to ensure the ordinary Adler zero. We find that the five-point scattering amplitude vanishes identically for this choice of couplings.

This bootstrap suggests that there should exist a solid with an enhanced Adler zero.  To investigate this possibility, we generalize to a solid in arbitrary spacetime dimension while including all $[B^n]$ operators in the Lagrangian.  Remarkably, we find that 
 the three-, four-, and five-point scattering amplitudes for a solid with the constraints in Eqs.~\eqref{eq:condsolids1}, ~\eqref{eq:condsolids2}, and \eqref{4pt5pt_bootstrap} agree with the scattering amplitudes derived from a physically equivalent Lagrangian, 
\eq{
      {\cal L}_{\rm exc.\; solid} = \frac{1}{\kappa}\sqrt{-\det\left(\eta_{\mu \nu}+ \kappa \left(\partial^\prime_\mu\pi_i\partial^\prime_\nu \pi^i \right)\right)}  \, ,
}{L_exceptional_solid}
with a coupling constant $\kappa=-(1-c^2)$, where we introduced $\partial^\prime_\mu$ such that $\partial^\prime_\mu\partial^{\prime\mu}=-\partial_t^2+c^2\partial_i^2$. In addition, we have verified that the theory in \Eq{L_exceptional_solid} yields a six-point amplitude that vanishes as ${\cal O}(q^2)$ in the soft limit.\footnote{Since \Eq{L_exceptional_solid} is expressed in terms of a determinant over spacetime indices, one must impose kinematics in a specific dimension in order to verify the enhanced soft theorem.  In this case, the on-shell identities in \App{sec:kinematics} must be supplemented with dimensionally specific Gram determinant constraints.}

Since the phonon indices in \Eq{L_exceptional_solid} are only contracted with each other, they effectively label an internal symmetry.  Moreover, \Eq{L_exceptional_solid} clearly describes a theory that linearly realizes the Lorentz symmetry with respect to the speed of sound $c$ in \Eq{eq:softboostNR}. Note that the coupling $\kappa$ vanishes when $c=1$, yielding a free theory. Thus, we have arrived at the Lagrangian for multiple relativistic DBI fields.  The enhanced soft limit we have encountered is not surprising in light of the results of \cite{Cheung:2014dqa, Cheung:2016drk,Cheung:2015ota}, which show that the only relativistic theory of single derivatively coupled scalar with an enhanced soft limit is DBI.

\subsubsection{Fluids}

Let us now attempt to construct an exceptional fluid theory. As noted earlier, we only consider the longitudinal external states. That makes the fluid case different from the solid case. The only condition for the fluid comes from requiring $A_{LLL}$ in \Eq{A3_solid} to vanish, subject to the fluid constraints in \Eq{lambda_fluid}. That fixes the free coupling in the fluid amplitude in terms of the speed of sound for the longitudinal modes, 
\eq{
\tau_1=-\tfrac{1}{16}(1-c_L^2)^2 \, .
}{eq:fluid3ptConstraint}
As for the solid, this is sufficient to ensure an Adler zero thanks to the soft theorem for spontaneously broken spatial translations in \Eq{soft_theorem_solid_translation}.
Similarly, the soft theorem for spontaneously broken boosts in \Eq{soft_theorem_solid_Lorentz} shows that an enhanced Adler zero requires an emergent relativistic symmetry.

Next we demand that the four-point amplitude for all longitudinal polarizations has an enhanced Adler zero. Imposing the constraints in Eqs.~\eqref{lambda_fluid} and \eqref{eq:fluid3ptConstraint}, we find that the four-point amplitude is
\eq{
A_{LLLL} =  -\frac{3}{c_L^4}\left( 5 (1 -c_L^2)^3-128 \tau_2 \right)\omega_1\omega_2\omega_3\omega_4+\frac{1-c_L^2}{4c_L^2} (s^{2}_{12}+s^{2}_{13}+s^{2}_{14}) \,  ,
}{eq:fluid4ptamp}
where $s_{ab}$ was defined in \Eq{eq:mandelstams} and here $c=c_L$. In the soft limit, the first term scales as $\mathcal O(q)$ and the second as $\mathcal O(q^2)$. Imposing the enhanced Adler zero constraints the coupling to be
\eq{
\tau_2 = \tfrac{5}{128} (1 - c_L^2)^3 \,.
}{eq:fluid4ptConstraint}
From the choice of coefficients in \Eqs{eq:fluid3ptConstraint,eq:fluid4ptConstraint} it is easy to guess the pattern for the exceptional fluid Lagrangian
\begin{align}
   {\cal L}_{\rm exc.\;fluid} 
   &= \frac1{1 - c_L^2}\sqrt{1+(1 - c_L^2)\text{det}B'} - \frac1{1 - c_L^2} \\
   &=  -\frac12 \text{det}B' +\frac1{8} (1 - c_L^2)\text{det}B'^2 -\frac1{16} (1 - c_L^2)^2 \text{det}B'^3 + \frac5{128} (1 - c_L^2)^3 \text{det}B'^4 + \cdots  \nonumber
    \,.
\end{align}
 Although not obvious, can we identify a different Lagrangian which reproduces exactly these fluid amplitudes which have an enhanced soft limit. That Lagrangian takes the form
 \eq{
   {\cal L}'_{\rm exc.\;fluid} = \frac{1}{\kappa^\prime}\sqrt{1+\kappa^\prime\left(\dot{\pi}^2-c_L^2(\partial_i \pi^i)^2\right)} \,.
}{}
The coupling constant is $\kappa^\prime = - (1-c_L^2)/c_L^2$. As it turns out, this Lagrangian is tree-level equivalent to relativistic DBI with the longitudinal phonon playing the role of the DBI scalar.  To understand why this is so, we simply expand the phonon field in terms of its longitudinal and transverse components as in \Eq{LT_decomp}, yielding
\eq{
   {\cal L}'_{\rm exc.\; fluid} = \frac{1}{\kappa^\prime}\sqrt{1+\kappa^\prime\left(\dot{\pi}_T^2+\dot{\pi}_L^2-c_L^2(\partial_i \pi_L^i)^2\right)} \, .
}{}
Crucially, since $\pi_T$ enters quadratically, it can only be pair-produced.  Thus, for tree-level amplitudes with all external legs with longitudinal polarizations, the transverse modes decouple completely.  Hence, the resulting Lagrangian is equivalent to that of DBI for the longitudinal phonon mode at tree level. 

\subsubsection{Superfluids}

A similar analysis was carried out for the superfluid in \cite{Pajer:2018egx,Green:2022slj}. In order to have an Adler zero, one demands that the tree-point amplitude in \Eq{A3_superfluid} vanishes, which imposes $g_3=0$. Then the soft theorem in \Eq{soft_theorem_superfluid_boost_otherbasis} shows that an enhanced Adler zero requires an emergent boost symmetry. From directly requiring the ${\cal O}(q^2)$ vanishing of the soft limits of higher-point amplitudes it follows that the only such theory is
\eq{
 {\cal L}_{\rm exc.\; superfluid} = \frac{1}{c}\sqrt{1+(1-c^2)\left(\partial_\mu\phi \partial^\mu\phi\right)}\, ,
}{}
written in terms of the field $\phi$ defined in \Eq{eq:superfluid_phi}. This superfluid DBI Lagrangian corresponds to one of the symmetric superfluid theories in \cite{Pajer:2018egx}, found by considering all possible new symmetries that form a consistent algebra with Poincar\'e and $U(1)$ shift symmetries.

As before, we can identify a classically equivalent Lagrangian 
\eq{
   {\cal L}'_{\rm exc.\; superfluid} = \frac{1}{\kappa''}\sqrt{1+\kappa''\left(\dot{\varphi}^2-c^2(\partial_i \varphi)^2\right)} \,,
}{}
with $\kappa'' = -(1-c^2)/c^4$, which also follows from the emergent boost symmetry and the results in Refs.~\cite{Cheung:2014dqa, Cheung:2016drk,Cheung:2015ota}. This Lagrangian describes a brane moving with constant velocity in an extra dimension \cite{Grall:2020ibl}.

A more general soft bootstrap for nonrelativistic theories with a single scalar was also performed in \cite{Mojahed:2022nrn}, including the case of NGB with quadratic dispersion relations. By imposing the enhanced Adler zero for a theory with a single scalar with one derivative per field---such as the superfluid---the resulting theory was also found to be effectively relativistic.

\subsection{Exceptional Framons}

The analysis of enhanced soft limits for framids is slightly different. Framon interactions, unlike phonons, do not involve one derivative per field. This is analogous to what happens in the relativistic NLSM, where at leading order in the derivative expansion the interactions have the structure $\pi^{n} (\partial_\mu \pi)^2$.

The framon soft theorem in  \Eq{soft_thm_framid} shows that an Adler zero requires  the vanishing of the three-point amplitude. This is achieved for
\eq{
c_T = c_L = 1\,,
}{eq:condframids1}
corresponding to a genuine relativistic dispersion relation for all modes. With this choice the soft theorem still yields
\eq{
&\lim_{q\rightarrow 0}   A^{i_1 \cdots i_n i}_{n+1}(p_1, \cdots, p_n, q)= -\sum_{a=1}^n \left(\omega_a \tfrac{\partial}{\partial \vv p_{ai}}+ \vv p_a^i \tfrac{\partial}{\partial\omega_a}\right)  \left[ A^{i_1 \cdots i_a \cdots i_n }_{n} (p_1, \cdots, p_a, \cdots, p_n)\right] \, ,
}{soft_thm_framid_onlyboost}
so an Adler zero requires full boost symmetry of the framon amplitudes. This is indeed a consequence of the choice in \Eq{eq:condframids1} which corresponds to the Lagrangian
\eq{
{\cal L}_{\rm exc.\; framid} =  -\tfrac12 M_2^2 (\partial_\mu A_\nu)^2\,,
}{}
at the leading order in the EFT derivative expansion.
Note that the derivative indices are independent from the Lorentz indices of $A_{\mu}$. Thus, this theory simply describes a relativistic NLSM realizing the spontaneous breaking of an internal $SU(2)$ symmetry corresponding to the boosts of $A_{\mu}$. That such choice of couplings at this order corresponds to the relativistic NLSM was already pointed out in \cite{Nicolis:2015sra}. Here we have derived this condition from the bottom up as a necessary condition for an Adler zero. Furthermore, since our soft theorem is a consequence of symmetry it is valid to all orders in the EFT derivative expansion.

\subsection{Alternative Bootstraps}
Given the soft theorem in \Eq{soft_theorem_exact}, can we extend the approach of on-shell soft bootstrap to theories with nonzero soft limits? Immediately, we see an obstruction: as we have noted, the soft theorem in \Eq{soft_theorem_exact} contains field basis dependent terms: the off-shell three-point vertex $V_3$ and the field transformation under the symmetry $\beta$.   If we do not know these beforehand, how can we initiate a bootstrap procedure?

Despite this apparent lack of data, such a bootstrap is actually possible in some circumstances, e.g., for the framid. Suppose we want to find theories with two derivatives per interaction vertex which spontaneously break Lorentz boosts. Let us start with the general soft theorem in \Eq{soft_theorem_exact}, together with \Eq{boost_framid} for the framons. For this particular case, the soft theorem for the on-shell three-point amplitude encodes enough information to constrain all couplings in $V_3$. Then, imposing the soft theorem on higher point amplitudes recursively, we obtain a set of constraints on the couplings in a Lagrangian ansatz with only rotational symmetry. We find that the only solution at the two-derivative order coincides with the expansion of the framid Lagrangian in \cite{Nicolis:2015sra}, which was constructed using the top-down approach. 

Similarly, higher-derivative deformations of the framid Lagrangian can be obtained either from the top-down or the bottom-up construction (by adding higher dimensional operators to the Lagrangian ansatz). The two methods are complementary, however the bottom-up on-shell approach allows us to bypass the field-dependent redundancies of the Lagrangian and directly obtain the on-shell framid amplitudes.

\section{Conclusions}
\label{sec:Conclusion}

In this paper we have initiated a systematic analysis of the soft behavior of scattering amplitudes in a broad class of condensed matter systems.  The common thread linking these theories is that their gapless modes are the NGBs of spontaneously broken spacetime symmetries.  As per the classification of \cite{Nicolis:2015sra}, the dynamics of superfluids, solids, fluids, and framids can all be derived from universal principles governing nonlinearly realized symmetries.

Using current conservation, we have derived the general soft theorem in \Eq{soft_theorem_exact}, which encodes the action of broken symmetry generators on the NGBs.  The ingredients entering the soft theorem are the parameters of the broken symmetry transformation, $\alpha$ and $\beta$, together with the propagator and cubic vertex of the theory, $\Delta$ and $V_3$.   While $\beta$ and $V_3$ are generally dependent on the field basis, and thus not individually invariant, they enter into \Eq{soft_theorem_exact} in a way that is field basis independent.  Furthermore,  \Eq{soft_theorem_exact} should be viewed as a soft theorem because it is an operation relating on-shell scattering amplitudes.

Applying this construction to various EFTs, we present and check a broad array of soft theorems, including those corresponding to temporal translations and Lorentz boosts of the superfluid, spatial translations and Lorentz boosts of the solid, volume-preserving diffeomorphisms of the fluid, and Lorentz boosts of the framid.

Last but not least, we have applied a soft bootstrap approach to these condensed matter systems.  In this analysis we take as input the assumption of an enhanced Adler zero condition for soft NGBs.  While we have identified exceptional theories of the solid and fluid with these enhanced infrared properties, they are all closely related in structure to relativistic DBI.

Our analysis leaves a number of directions for future work.  For example, as noted earlier, our soft theorem does not actually require $SO(3)$ rotation invariance in the broken phase.  Indeed, the only requirement is that symmetry breaking preserves some notion of a conserved energy and momentum. For this reason one can in principle study condensed matter systems with even less rotational symmetry.  It would be interesting to classify these theories and their corresponding soft theorems.

Throughout this paper we have focused on scattering induced by the self-interactions of phonons.  However, the interactions of phonons with other degrees of freedom, e.g., crystal defects or vortices are also of interest for many condensed matter systems.  As long as these other modes can be incorporated consistently into the EFT of spontaneous spacetime symmetry breaking, it should be possible to mechanically derive new soft theorems for scattering processes involving these other degrees of freedom.

Another avenue for exploration is more elaborate variations of the soft bootstrap.  In particular, it should be possible to {\it assume} a general ansatz for the broken symmetry parameters $\alpha$ and $\beta$, as well as a general ansatz for the propagator and cubic vertex, $\Delta$ and $V_3$.  Sculpting out the space of amplitudes satisfying our soft theorem might offer insight into new condensed matter systems of interest.

Finally, it would be interesting to understand whether the geometric perspective on soft theorems presented in \cite{Cheung:2021yog, Cheung:2022vnd} extends to a nonrelativistic setting. A geometric description of the corresponding EFTs has already been described in \cite{Watanabe:2012hr}, so generalizing to this case should be relatively straightforward.

\begin{center} 
   {\bf Acknowledgments}
\end{center}
\noindent 
We are grateful to Tomas Brauner, James Mangan, Ira Rothstein, and Chia-Hsien Shen for useful discussions and comments on the paper.
This work is supported by the DOE under grant no.~DE-SC0011632 and by the Walter Burke Institute for Theoretical Physics. J.P.-M. is also supported in part by the NSF under Grant No. NSF PHY-1748958 and would like to thank Institut des Hautes \'{E}tudes Scientifiques, the Korea Institute for Advanced Study, and the Kavli Institute for Theoretical Physics for hospitality while this work was being completed.

\appendix

\section{Nonrelativistic Kinematics}
\label{sec:kinematics}

Our analysis will require a careful treatment of on-shell kinematics for scattering amplitudes with nonrelativistic dynamics.  For the $n$-point amplitude, $A_n$, we define the four-momenta of the $n$ hard legs to be
\eq{
p_{a}^{\mu} = (\omega_a, \vv p_a^i) \, ,
}{}
where the external particle index is an integer in the range $1 \leq a \leq n$.

It will be crucial to define the notion of a minimal on-shell basis of kinematic invariants.  A priori, the $n$-point amplitude is an $SO(3)$ invariant quantity that is a function of the energies $\omega_a$ and all inner products of three-momenta, $\vv p_a \cdot \vv p_b$ for $1\leq a\leq b\leq n$.    A minimal on-shell basis defines a set of kinematic variables for which all on-shell constraints are automatically imposed.

To achieve this, we first eliminate the energy and three-momentum of some leg, chosen here to be leg $n$, so
\eq{
\omega_n = -\sum_{a=1}^{n-1} \omega_a \qquad \textrm{and} \qquad \vv p_n^i =  -\sum_{a=1}^{n-1} p_a^i \, .
}{}
The elimination of the energy $\omega_n$ and three-momentum $\vv p_n^i$ of leg $n$ via the above equations then automatically enforces total momentum conservation.  Second, we impose the on-shell conditions for the external legs, allowing us to eliminate the kinematic invariants
\eq{
 \vv p_{a}^2 = \frac{\omega_a^2}{ c_a^2} \qquad \textrm{for} \qquad 1\leq a \leq n-1 \,,
}{}
where $c_a$ is the speed of sound for the corresponding leg.  For example, for phonons it would be the longitudinal or transverse speeds of sound, $c_L$ or $c_T$.  The above condition allows us to eliminate $\vv p_a^2$ for $1\leq a \leq  n-1$. However since we have already eliminated $\vv p_n^i$ by momentum conservation, for the case of $a=n$ the on-shell condition imposes a more elaborate constraint,
\eq{
 \vv p_n^2 = \left(\sum_{a=1}^{n-1} \vv p_a \right)^2 =\frac{1}{c_n^2}\left(\sum_{a=1}^{n-1} \omega_a \right)^2 \,,
}{leg_n_onshell}
which should can be used to eliminate one more kinematic invariant, which we can choose to be $\vv p_{n-2}\cdot\vv  p_{n-1}$ without loss of generality.
In summary, the minimal kinematic basis is comprised of the variables
\eq{
\omega_a &\qquad \textrm{for} \qquad 1\leq a \leq n-1 \,, \\
 \vv p_a\cdot \vv p_b &\qquad \textrm{for} \qquad  1 \leq a < b \leq n-2  \qquad\textrm{and} \qquad 1 \leq a \leq  n-3, \;\; b=n-1 \,,
}{min_kin}
with all other kinematic variables fixed by on-shell conditions.

With the inclusion of external polarization vectors,  $\vv e_a^i$ for $1\leq a\leq n$, similar logic applies.  Without assuming any special properties of the external polarizations, for example whether they are longitudinal or transverse, we simply eliminate all invariants involving $\vv p_n^i$.  Hence, we obtain
\eq{
 \vv e_a\cdot \vv e_b &\qquad \textrm{for} \qquad  1 \leq a < b \leq n \,, \\
 \vv p_a\cdot \vv e_b &\qquad \textrm{for} \qquad  1 \leq a  \leq n-1, \;\; 1\leq b\leq n \,,
}{min_kin_pol}
for the elements of the minimal kinematic basis involving polarizations.

Let us also write down the explicit minimal kinematic basis for three-point scattering,
\eq{
\textrm{ basis for } A_3:\qquad &\omega_1, \,\; \omega_2\, , \\
& \vv p_1 \cdot \vv e_1 ,\,\; \vv p_1 \cdot \vv e_2 ,\,\; \vv p_1 \cdot \vv e_3 ,\,\;  \vv p_2 \cdot \vv e_1 ,\,\; \vv  p_2 \cdot \vv e_2 ,\,\; \vv p_2 \cdot \vv e_3 \, ,  \\
& \vv e_1 \cdot \vv e_2 ,\,\; \vv e_1 \cdot \vv e_3 ,\,\; \vv e_2 \cdot \vv e_3  \,  .
}{min_kin_3} 
 The utility of these variables is that we can freely change them while remaining on the kinematic surface that defines on-shell configurations.  From here on, we will write all on-shell quantities in terms of these minimal bases.
 
Finally, to evaluate the amplitude for a specific configuration of external modes, we plug in explicit longitudinal or transverse polarizations. The trasverse conditions put additional constraints on the minimal basis
\eq{
\vv p_a\cdot \vv e_a &= 0 \qquad \textrm{for} \qquad  1 \leq a < n \,, \\
\vv p_{n-1} \cdot \vv e_{n} &= - \sum_{a=1}^{n-2} \vv p_a \cdot \vv e_n  \, .
}{min_kin_polT}

 For our analysis, we will be interested in how the soft limit of the $(n+1)$-point amplitude, $A_{n+1}$, can be expressed in terms of the $n$-point amplitude, $A_n$.  For this reason, we define legs $1, \cdots, n$ to be hard, since they are present in both $A_{n+1}$ and $A_n$. On the other hand, leg $n+1$, with external polarization $e^{i}$, will be taken soft, so we parameterize it with a special four-momentum
\eq{
 q^\mu=(\omega,\vv q^i)  \, .
}{}
To be very explicit, $A_{n+1}$ is a function of $p_1, \cdots, p_n, q$ while $A_n$ is a function of $\vv , \cdots, p_n$.  Both should be evaluated in a minimal kinematic basis in which the energy $\omega_n$, the three-momentum $\vv p_n^i$, and the invariant $\vv p_{n-2}\cdot \vv p_{n-1}$ have been eliminated.    Consequently, for {\it any values} of the soft momentum $q$, the amplitudes remain on-shell.  This ensures that the soft limit is taken while maintaining all on-shell conditions.  The minimal basis for four-point scattering is then
\eq{
\textrm{ basis for } A_4:\qquad &\omega, \,\; \omega_1, \,\; \omega_2 , \,\; \vv q \cdot \vv p_1, \,\; \vv q\cdot \vv p_2 , \,\;  \vv q \cdot \vv e ,\,\; \vv q \cdot \vv e_1 ,\,\; \vv q \cdot \vv e_2 ,\,\;  \vv q \cdot\vv  e_3 \, ,    \\
& \vv p_1 \cdot \vv e ,\,\; \vv p_1 \cdot \vv e_1 ,\,\; \vv p_1 \cdot \vv e_2 ,\,\; \vv  p_1 \cdot \vv e_3,\,\; \vv p_2 \cdot \vv e ,\,\; \vv p_2 \cdot \vv e_1 ,\,\; \vv p_2 \cdot \vv e_2 ,\,\;  \vv p_2 \cdot \vv e_3  \, ,\\
& \vv e \cdot \vv e_1 ,\,\; \vv e \cdot \vv e_2 ,\,\; \vv e \cdot \vv e_3 ,\,\;  \vv e_1 \cdot \vv e_2 ,\,\; \vv  e_1 \cdot \vv e_3 ,\,\; \vv e_2 \cdot \vv e_3  \,  .
}{min_kin_4}
By construction, the minimal basis for $A_4$ in \Eq{min_kin_4} reduces to the minimal basis for $A_3$ in \Eq{min_kin_3} in the soft limit, $q\rightarrow 0$.

While the above approach is somewhat convoluted, we emphasize that any definition of the soft limit of an on-shell amplitude requires something analogous.  In general, simply changing the momentum $q$ of a leg to be soft will not maintain on-shell conditions.  For the case of on-shell soft recursion \cite{Cheung:2015ota, Cachazo:2016njl, Luo:2015tat, Mojahed:2021sxy}, the soft limit of a given leg must always be compensated by a slight shift of one of the hard legs.  The minimal basis construction we have described above achieves this automatically.

\newpage
\bibliographystyle{utphys-modified}
\bibliography{bibl}
\end{document}